\documentclass[a4paper, pre, twocolumn,superscriptaddress,floatfix,amssymb,showpacs,longbibliography]{revtex4-1}
\pdfoutput=1
\usepackage[pdfpagemode=None,colorlinks=true,urlcolor=black,%
linkcolor=blue,citecolor=blue,pdfstartview=FitH]{hyperref}
\usepackage{enumerate}
\usepackage{times}
\usepackage{graphicx}
\usepackage{amsmath}
\usepackage{color}
\usepackage{natbib}

\graphicspath{{fig/}} 
\begin{document}

\title{Uniform description of polymer ejection dynamics from capsid with and without hydrodynamics}

\author{J. Piili}
\author{P. M. Suhonen}
\author{R. P. Linna}
\email{Author to whom correspondence should be addressed: riku.linna@aalto.fi}
\affiliation{Department of Computer Science, Aalto University, P.O. Box 15400, FI-00076 Aalto, Finland}

\begin{abstract}
We use stochastic rotation dynamics to examine the dynamics of the ejection of an initially strongly confined flexible polymer from a spherical capsid with and without hydrodynamics. The results obtained using SRD are compared to similar Langevin simulations. Inclusion of hydrodynamic modes speeds up the ejection but also allows the part of the polymer outside the capsid to expand closer to equilibrium. This shows as higher values of radius of gyration when hydrodynamics are enabled. By examining the waiting times of individual polymer beads we find that the waiting time $t_w$ grows with the number of ejected monomers $s$ as a sum of two exponents. When $\approx 63 \%$ of the polymer has ejected the ejection enters the regime of slower dynamics. The functional form of $t_w$ vs $s$ is universal for all ejection processes starting from the same initial monomer densities. Inclusion of hydrodynamics only reduces its magnitude. Consequently, we define a universal scaling function $h$ such that the cumulative waiting time $t = N_0 h(s/N_0)$ for large $N_0$. Our unprecedently precise measurements of force indicate that this form for $t_w(s)$ originates from the corresponding force towards the pore decreasing super exponentially at the end of the ejection. Our measured $t_w(s)$ explains the apparent superlinear scaling of the ejection time with the polymer length for short polymers. However, for asymptotically long polymers $t_w(s)$ predicts linear scaling.
\end{abstract}


\maketitle
\section{Introduction}\label{sec:intro}

Packaging and ejection of macromolecules in confinements are of high interest due to the potential technological and medical applications, such as drug delivery and gene therapy~\cite{glasgow}. The basic understanding of these processes is important also due to the relevant fundamental biological processes, the most prominent of which is the viral packaging in and ejection from bacteriophages~\cite{muthukumar1,muthukumar2,smith,grayson,ali,ghosal,cacciuto2,sakaue_polymer_decompression,riku_dynamics_of_ejection}. The theoretical treatments of polymer ejection are strictly based on fully flexible chains~\cite{muthukumar1,cacciuto2,sakaue_polymer_decompression}, in spite of the experimental studies being almost solely done on semiflexible double-stranded DNA. This makes sense, since theoretically ejection from confinements is most intriguing when the spring force of the semiflexible polymer does not dominate over the more subtle mechanisms. Besides, the ejection of fully flexible polymers is highly relevant outside the purely theoretical realm due to many important polymers, such as proteins, single-stranded DNA, and RNA, belonging to this class.

In our previous study we showed that the blob-scaling picture used as a basis for analyzing the ejection dynamics from strong confinement is not valid. The blob picture presumes semidilute conditions, which does not hold for {\it in vivo} encapsulated polymers. In computer simulations polymers are far too short to justify the blob-scaling assumption. In spite of these shortcomings the blob-scaling has been used to explain the apparent scaling of the ejection time $\tau$ with the length of the polymers $N_0$. Indeed, if only the ejection time $\tau$ as a function of $N_0$ is measured, $\tau$ seemingly scales superlinearly with $N_0$. However, by inspecting the waiting times $t_w(s)$ we showed that there, in fact, is no such scaling. $t_w(s)$ is defined as the time it takes for a monomer labeled $s$ to translocate after the previous monomer $s-1$ has translocated. From our simulations using a hybrid method consisting of molecular dynamics (MD) and stochastic rotation dynamics (SRD)~\cite{malevanets_orig,malevanets_mesoscopic} we obtained $t_w$ that grows essentially exponentially with $s$~\cite{piili_capsid}.

Due to this exponential growth of $t_w(s)$ conflicting the theoretical predictions, some suspicion was cast on the simulation method. Accordingly, to be conclusive a confirmation using a well established method is called for. To this end, we have implemented an identical  capsid model in our Langevin Dynamics (LD) algorithm~\cite{allen}. LD algorithm is a numerical implementation of a stochastic differential equation describing Brownian motion of particles, so it can be regarded as the most fundamental method available for a dynamical simulation of polymer ejection. We present here a thorough comparison of polymer ejection dynamics obtained using LD and SRD.

Our main objective in the present paper is to establish precise forms for $t_w(s)$ in the absence and presence of hydrodynamic interactions, that is, to determine how the inclusion of hydrodynamic modes changes the ejection dynamics. To achieve this we use SRD where hydrodynamic modes can easily be switched on or off. Measuring $t_w(s)$  is the most precise way of gaining detailed information on ejection dynamics. The form of $t_w(s)$ reflects the form of the force $f$ reduced to the pore during the ejection. We determine $f(s)$ with high precision using LD that due to being computationally more effective than SRD allows us to gain much better statistics than what was possible in our previous study using only SRD~\cite{piili_capsid}. This study reveals the surprisingly strong influence of the local effects in the vicinity of the pore on the overall ejection dynamics. Indications of this were observed already in our earlier study on capsid ejection, where a force applied to aligning a polymer close to the pore was found to give an effective bias to the ejection~\cite{riku_dynamics_of_ejection}.

The paper is organized as follows. The computational models are described in Section~\ref{models}. The procedure for matching the models based on LD and SRD is described at the end of this section. Results are presented in Section~\ref{sec:res}. In this section we first compare the results obtained by SRD and LD after which we extract precise forms for $t_w(s)$ from each model thus pinpointing the effect of hydrodynamics. We then present our measurements of the radius of gyration of the polymer segment outside the capsid and the force at the pore. By analyzing these measurements we are able to give an accurate account of the prevailing mechanisms during polymer ejection. Finally, in Section~\ref{sec:con} we summarize our results and present the conclusions based on them.

\begin{figure}
\includegraphics[width=\linewidth]{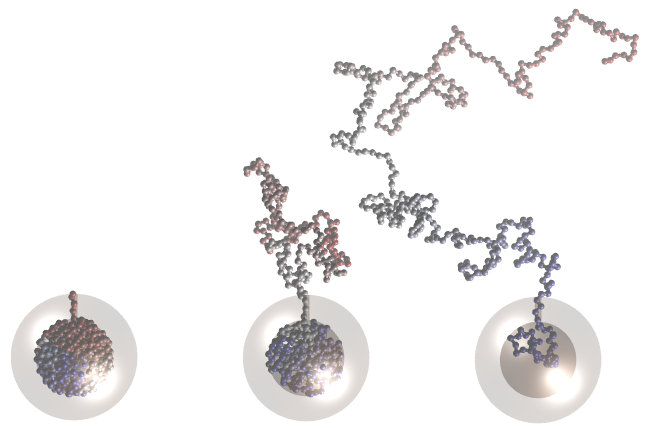}
\caption{(Color online) Snapshots of an ejection simulation at different times. The polymer ejects from a solid spherical capsid through a narrow pore. (Image created using VMD~\cite{vmd} and POV-Ray~\cite{povray})}
\label{fig:capsidPicture}
\end{figure}

\section{The Computational Models}
\label{models}

Here, we describe the computational models. The primary computational method is stochastic rotation dynamics (SRD), also called multi-particle collision dynamics, that allows for the inclusion of hydrodynamics~\cite{malevanets_orig,malevanets_mesoscopic}. We validate our SRD model by closely comparing it to the identically implemented model, see Fig.~\ref{fig:capsidPicture}, in our Langevin dynamics (LD) algorithm. LD is based on a thoroughly analyzed and understood stochastic differential equation and hence serves as a perfect reference for our SRD model in the case where hydrodynamics is switched off~\cite{allen}. LD also has the benefit of being computationally much more efficient than SRD.

\subsection{The polymer model}
Polymers are modeled as chains of point-like beads of mass $m_b$. Adjacent beads are connected via the Finitely Extensible Nonlinear Elastic (FENE) potential
\begin{align}\label{equ:fene}
U_F = - \frac{K}{2} r_{\rm max}^2 \ln{\left( 1 - \left(\frac{r}{r_{\rm max}} \right)^2 \right)}\;,\; r < r_{\rm max},
\end{align}
where $r$ is the distance between adjacent beads and $K$ and $r_{\rm max}$ are potential parameters describing the strength and maximum  distance limit of adjacent beads. Each bead interacts with all other beads via the (shifted and truncated) Lennard-Jones potential
\begin{align}\label{equ:lj}
U_{LJ} = \left\lbrace \begin{array}{ccl}
		4.8 \epsilon \left[\left(\frac{\sigma}{r_{ij}}\right)^{12} - \left(\frac{\sigma}{r_{ij}}\right)^6\ \right] + 1.2\epsilon &,& r_{ij} \leq \sqrt[6]{2}\sigma\\
		0		&,& r_{ij} > \sqrt[6]{2}\sigma
		\end{array},
 \right.
\end{align}
where $\epsilon$ and $\sigma$ are potential parameters and $r_{ij}$ is the distance between beads $i$ and $j$. The potential is truncated at $r = \sqrt[6]{2}\sigma$ in order to model a good solvent. The potential parameters are chosen as $\sigma = 1.0$, $\epsilon = 1.0$, $K = 30/\sigma^2$, and $r_{\rm max} = 1.5\sigma$ in reduced units.

\subsection{The solvent and polymer dynamics}
The polymer is immersed in a solvent modeled by stochastic rotation dynamics (SRD)~\cite{malevanets_orig,malevanets_mesoscopic}. The SRD method was chosen because it allows taking both hydrodynamics and Brownian motion directly into account in a computationally feasible way. A particular benefit of the method is the possibility to switch off hydrodynamics to better understand its effects. This also allows us to verify the polymer escape in SRD against that in Langevin dynamics.

The SRD solvent consists of point-like particles whose dynamics can be divided into the \textit{streaming} and the \textit{collision} steps. In the streaming step the solvent particles are moved ballistically,
\begin{align}
\mathbf{r}_i(t + \Delta t) = \mathbf{r}_i(t) + \mathbf{v}_i(t)\Delta t,
\end{align}
where $t$ is the simulation time, $\Delta t$ is the SRD time step, $\mathbf{r}_i$ is the position, and $\mathbf{v}_i$ is the velocity of solvent particle $i$. If in this step the solvent particle hits the capsid wall, it is bounced back to the direction of incidence and its velocity is reversed. In other words, the capsid wall constitutes a no-slip boundary for the particle. No-slip boundary conditions ensure that the flow velocity in the surface of a wall is zero~\cite{lamura_srd_poseuille}.

In the collision step the simulation space is divided into a grid of cubic cells whose edges are of length $1.0$. The interactions between particles are modeled by rotating the random part of particle velocities within each cell by the equation
\begin{align}\label{equ:srd_step}
\mathbf{v}_i(t+\Delta t) = \mathbf{v}_{\rm cm}(t) + \Omega \left[ \mathbf{v}_i(t) - \mathbf{v}_{\rm cm}(t) \right],
\end{align}
where $\mathbf{v}_{\rm cm}(t)$ is the center-of-mass velocity of the particles in the cell and $\Omega$ is a rotation of angle $\theta$ around a randomly chosen axis. The rotation axis is drawn randomly for each cell each time step. The rotation angle is chosen as $\theta = 3\pi/4$. It can be used to adjust the viscosity of the solvent. The solvent is kept at constant temperature of $k T = 1.0$ by scaling the random part of particle velocities such that the equipartition theorem holds at each time step~\cite{frenkel_moldy}. The density of the SRD solvent was chosen such that on average there are 5 particles per unit volume. When hydrodynamics is included $\eta = 4.67$ is obtained for the solvent viscosity in reduced units with the chosen parameter values~\cite{kikuchi_transport}.

The polymer is coupled with the solvent in the collision step where the velocities of polymer beads are updated similarly as those of solvent particles, see Eq.~\eqref{equ:srd_step}. The collisions retain the total momentum and energy within each cell. In order to maintain Galilean invariance, the grid is shifted randomly at each time step~\cite{ihle_galilean}. Hydrodynamic interactions can be switched off by randomly permuting the solvent particles' velocities after each collision step.

The polymer performs molecular dynamics (MD). In the velocity Verlet algorithm used for polymer dynamics, the time step is chosen as $\delta t = 0.0002$. The SRD time step $\Delta t = 0.5$. A relatively small $\delta t$ was used because with larger time steps the numerical errors accumulate inside the capsid when the polymer is tightly packed. The MD and SRD steps are performed in turns such that after $\Delta t/\delta t = 2500$ velocity Verlet steps a single SRD step is performed (including polymer in collision step of Eq.~\eqref{equ:srd_step}). The mass of the polymer beads $m_b = 16$ and the mass of the SRD particles $m_s=4$.

\subsection{The simulation geometry and initial polymer conformations}

The simulation geometry is depicted in Fig.~\ref{fig:capsidPicture}. A polymer ejects from inside a spherical capsid shell through a narrow pore of radius 0.4. The inside of the capsid is referred to as the \textit{cis} side and the outside of the capsid is referred to as the \textit{trans} side. The thickness of the capsid shell is 3. The radius $R_0$ of the inner shell of the capsid depends on the chosen initial monomer density $\rho_0$ and the initial number of polymer beads $N_0$ inside the capsid via
\begin{align}
 \rho_0 = \frac{N_0}{\frac43 \pi R_0^3}.
\end{align}
Notice that in some publications volume fraction $\phi_0 = 4/3\pi\rho_0$ is used, instead. Also the beads have often a hard sphere potential. In simulations using molecular dynamics, as in the present study, soft sphere potentials must be used. Hence, the values are not directly comparable. In effect, the largest densities used here supersede the densities used in most earlier studies (see ~\cite{piili_capsid}).

The capsid geometry is created using constructive solid geometry technique~\cite{wyvill_csg}, which we have implemented for use with the SRD and LD.
In the method intersections with the polymer particles' trajectories and capsid walls are traced and collisions are handled by slip boundary conditions. As stated before, for the solvent particles no-slip boundary conditions are applied, instead. We use a pore of radius 0.8 for the solvent particles which is twice as wide as that for the polymer beads. The larger pore for the solvent allows for a smoother fluid flow in the pore while the narrow pore for the polymer prevents hairpinning.

The initial conformations are created by injecting polymers inside the capsid through the pore with a large enough packing force within the pore.  Force is ramped up until the polymer is packed. Different conformations result from using different initial random generator seeds. Creating an initial conformation also includes thermalizing the polymer by scaling the bead velocities so as to have the polymer reside at temperature $kT = 1$. Before ejection, a new SRD solvent is initialized for the created polymer conformation and the polymer is allowed to equilibrate for 2000 time units before the ejection is allowed to start. This way we create an ensemble of random initial conformations. Inevitably, conformations created this way may include knots that have been shown to affect the ejection rate~\cite{matthews_knot_ejection, marenduzzo_topological_dna_ejection}. Identification of knots is beyond the scope of the present study.

To estimate the effect of initial conformations on ejection dynamics we performed simulations where polymers having a bending potential~\cite{bending_potential} with persistence length of $\sim 20$ were packed. This resulted in fundamentally different, spooled conformations. After this we removed the bending potential, which made the chains flexible and released them for ejection. The ejection times measured and averaged over multiple ejections starting from random and spool conformations differed only slightly. This indicates that in our simulations the details of initial conformations do not significantly affect ejection dynamics. It is in place to point out that the packing method, for example whether allowing for intermittent relaxation or not, does have some effect on initial conformations and potentially to ejection dynamics. To our knowledge these effects have not been thoroughly investigated. The packing method we use here falls into the category of generally used methods and consequently in this respect our study is directly comparable to previous studies on polymer ejection.

\subsection{Matching the LD and SRD models via friction}\label{sec:langevinMapping}

Our LD algorithm is implemented as derived by Ermak~\cite{ermak}. LD is a stochastic method where solvent particles are not explicitly simulated but the polymer resides in a Brownian heath bath satisfying the Langevin equation
\begin{align}\label{equ:langevin}
\frac{{\rm d}\mathbf{p}_i}{{\rm d} t}(t) = -\xi \mathbf{p}_i(t) + \boldsymbol{\eta}_i(t) + \mathbf{f}_i(t),
\end{align}
where $\xi$, $\mathbf{p}_i(t)$, and $\boldsymbol{\eta}_i(t)$ are the friction constant, momentum and random force of the bead $i$, respectively. $\mathbf{f}_i(t)$ is the sum of all forces exerted on the bead $i$. $\boldsymbol{\eta}_i(t)$ is a zero mean delta correlated Gaussian process, with $\left\langle \boldsymbol{\eta}_i(t) \cdot \boldsymbol{\eta}_i(t') \right\rangle = 2\xi kT m_b \delta(t-t')$.

To make the LD simulations comparable with the SRD simulations, the polymer model, potential parameters, simulation geometry, and simulation temperature were chosen the same. However, LD allowed us to use a larger time step $\delta t = 0.001$ than SRD due to it being a more efficient thermostat. Only the friction parameter $\xi$ in LD does not have a direct mapping to the friction in SRD. In order to choose an appropriate value for $\xi$ in the LD capsid ejection model, we performed two straightforward simulations for different values of $\xi$.

\begin{figure}
\includegraphics[width=.7\linewidth]{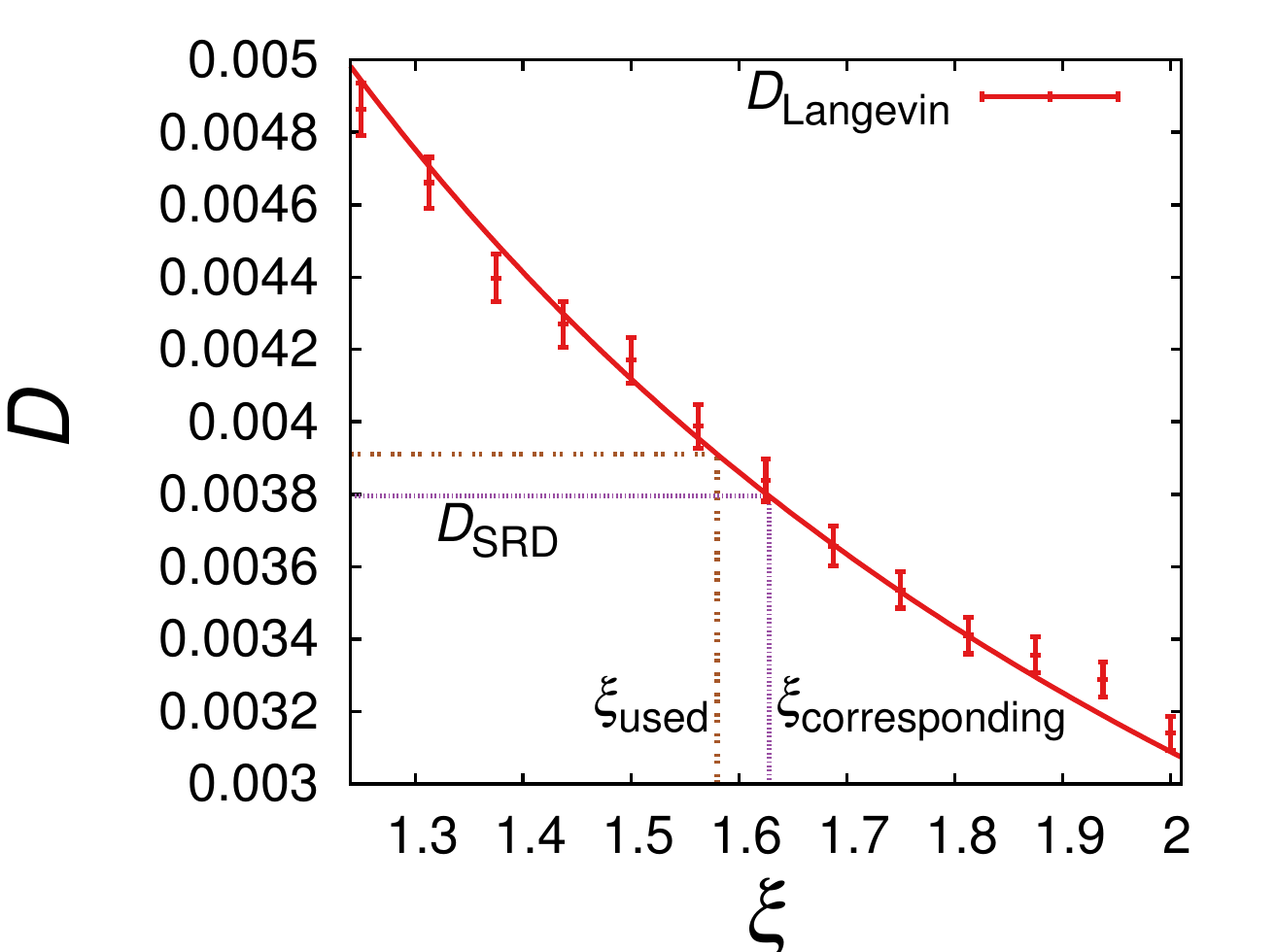}
\caption{(Color online) Diffusion constants $D$ of free polymers measured for different friction $\xi$ in LD simulations. The polymer length $N_0=10$. For SRD without hydrodynamics we measured a diffusion constant of $D_{\rm SRD} = 0.003796 \pm 0.000057$ that corresponds to the friction constant value $\xi = 1.63$ (depicted by purple dashed line). The friction was set at $\xi=1.58$ in the LD capsid ejection simulations (depicted by double dashed brown line). This value was chosen as it was found to slightly speed up the ejection times to better correspond to those of SRD without hydrodynamics.}
\label{fig:langevinMappingD}
\end{figure}

Figure~\ref{fig:langevinMappingD} shows the measured diffusion constant of a polymer of length $N_0=10$ in free solvent. Theoretically, the diffusion constant $D = k T / (N_0 m_b \xi)$~\cite{doi_introduction_to_polymer_physics}, which is close to the value in our LD simulations. $D$ in LD and SRD were found to coincide when $\xi = 1.63$. This is depicted by the dashed line in Fig~\ref{fig:langevinMappingD}. The double dashed line depicts the value of $\xi = 1.58$ that we chose for the capsid ejection simulations. The slightly smaller value for $\xi$ was chosen because it resulted in a better correspondence of the total capsid ejection times $\tau$ from simulations using LD and SRD without hydrodynamics. In other words, for some reason $\tau$ were found to be larger for LD for the same values of friction parameter $\xi$. Even with the choice $\xi = 1.58$ the ejection times are consistently larger in LD simulations.

\begin{figure}
\includegraphics[width=.745\linewidth]{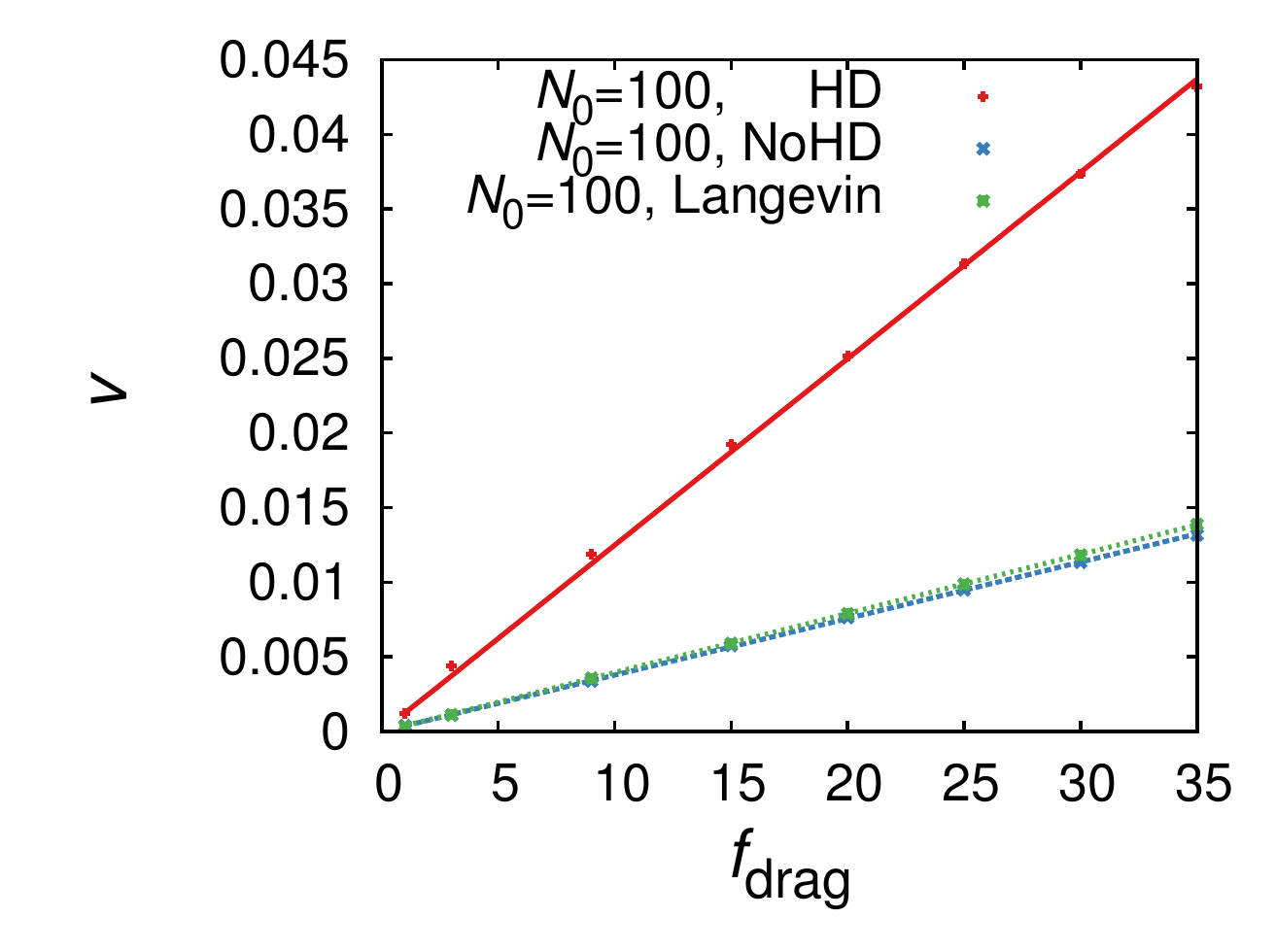}
\caption{(Color online) Measured terminal velocities $v$ of polymers of length $N_0=100$ dragged at an end bead by a constant force $f_{\rm drag}$. In simulations using SRD without hydrodynamics and LD the terminal velocities follow $v = f_{\rm drag}/(\xi m_b N_0)$ quite accurately. LD simulations were performed for $\xi = 1.58$. According to this measurement, the SRD simulation without hydrodynamics corresponds to $\xi = 1.65$. ($\xi \approx 0.50$ when hydrodynamics is included in SRD).}
\label{fig:langevinMappingV}
\end{figure}

To gain more confidence in the proper mapping of the correspondence between Langevin simulations and SRD without hydrodynamics, we performed simulations where a polymer of length $N_0=100$ is dragged by a constant force $f_{\rm drag}$ at the end bead. In the absence of hydrodynamics we expect the terminal velocity to follow
\begin{align}\label{equ:velocityVsForce}
 v = \frac{f_{\rm drag}}{\xi m_b N_0},
\end{align}
which can be obtained from Eq.~(\ref{equ:langevin}) by averaging over a long time, summing over all beads, and assuming that $\frac{{\rm d}\mathbf{p}_i}{{\rm d} t}(t)=0$ at terminal velocity. Figure~\ref{fig:langevinMappingV} shows $v$ measured for different $f_{\rm drag}$. In these LD simulations $\xi = 1.58$. By fitting we obtain the values $\xi_{\rm SRD,\;noHD} = 1.65$ and $\xi_{\rm SRD,\;HD} \approx 0.50$ for SRD without and with hydrodynamics, respectively. Observe, however that $\xi_{\rm SRD,\;HD}$ is not well defined for SRD with hydrodynamics since Eq.~\eqref{equ:velocityVsForce} is not accurate when hydrodynamics is included. Nevertheless, it is a good estimate on the effect the hydrodynamics has on the effective viscosity. The diffusion constant and terminal velocity measurements show that Langevin and SRD without hydrodynamics are in reasonable accordance for these basic systems.

\subsection{On polymer lengths and capsid volumes}
In this section we comment on relating the length and timescales in the simulations to the corresponding real-world scales. The fairly generic model in the present study is appropriate for validating the SRD method for simulating capsid ejection model and for characterizing the effect of hydrodynamics.

Since we use the fully flexible chain model, the persistence length of the polymer is $\lambda_p = \frac12 b$, where $b$ is the polymer segment length. In our simulations the equilibrium distance between consequent beads is $0.97$, which we take as the (average) segment length. $\lambda_p$ is of the order of $4$ nm for ssDNA~\cite{tinland_persistence_length} and $50$ nm for dsDNA~\cite{manning2006persistence}. Consequently, if we were to model ssDNA, one simulation unit would correspond to about $8$ nm. For dsDNA a simulation unit would correspond to about $100$ nm.

In our simulations, the capsid radii vary from $1.6$ ($N = 25,$ $\rho_0 = 1.5$) to $5.1$ ($N = 283,$ $\rho_0 = 0.5$). Thus, the persistence length is always an order of magnitude smaller than the capsid radius. For ssDNA the capsid radii would correspond to a range from 12.8 to 40.8 nm. For dsDNA the corresponding range is from 160 to 510 nm. A polymer of length $N_0 = 200$ would correspond to a ssDNA of length 1600 nm having about 4324 bases, since a single base is about 0.37 nm long~\cite{rechendorff_length_per_bead}. For dsDNA a polymer of length $N_0 = 200$ would correspond to a strand of length 20 000 nm with 59 000 base pairs (using 0.34 nm/bp).

If we expect for ssDNA $w \approx 1$ nm, then $\lambda_p \approx 8 w$. dsDNA width $w$ is about $2$ nm. So, for dsDNA $\lambda_p \approx 25 w$. In our simulations the repulsive LJ potential has an interaction distance of $\sim 1$, which can be taken as the approximate width of our flexible chain. Hence, $\lambda_p \approx 0.5 w$, falling short even for the ssDNA that the flexible chain in principle models. For maximum correspondence of coarse-grained polymers modeling real-world polymers in confinements the ratio $\lambda_p/w$ has to be adjusted via bending potential. We will look into this more closely in a forthcoming paper.

The estimated volume fraction inside the bacteriophage lambda is
\begin{align}
 \frac{V_{\rm DNA}}{V_{\rm capsid}} = \frac{48 502 \;\pi (1 \;{\rm nm})^2 0.34 \;{\rm nm}}{\frac43\pi (27.5 \;{\rm nm})^3} \approx 0.6.
\end{align}
This is in the same range as volume fractions in our simulations.

\begin{figure*}[!ht]
\includegraphics[width=.33\linewidth]{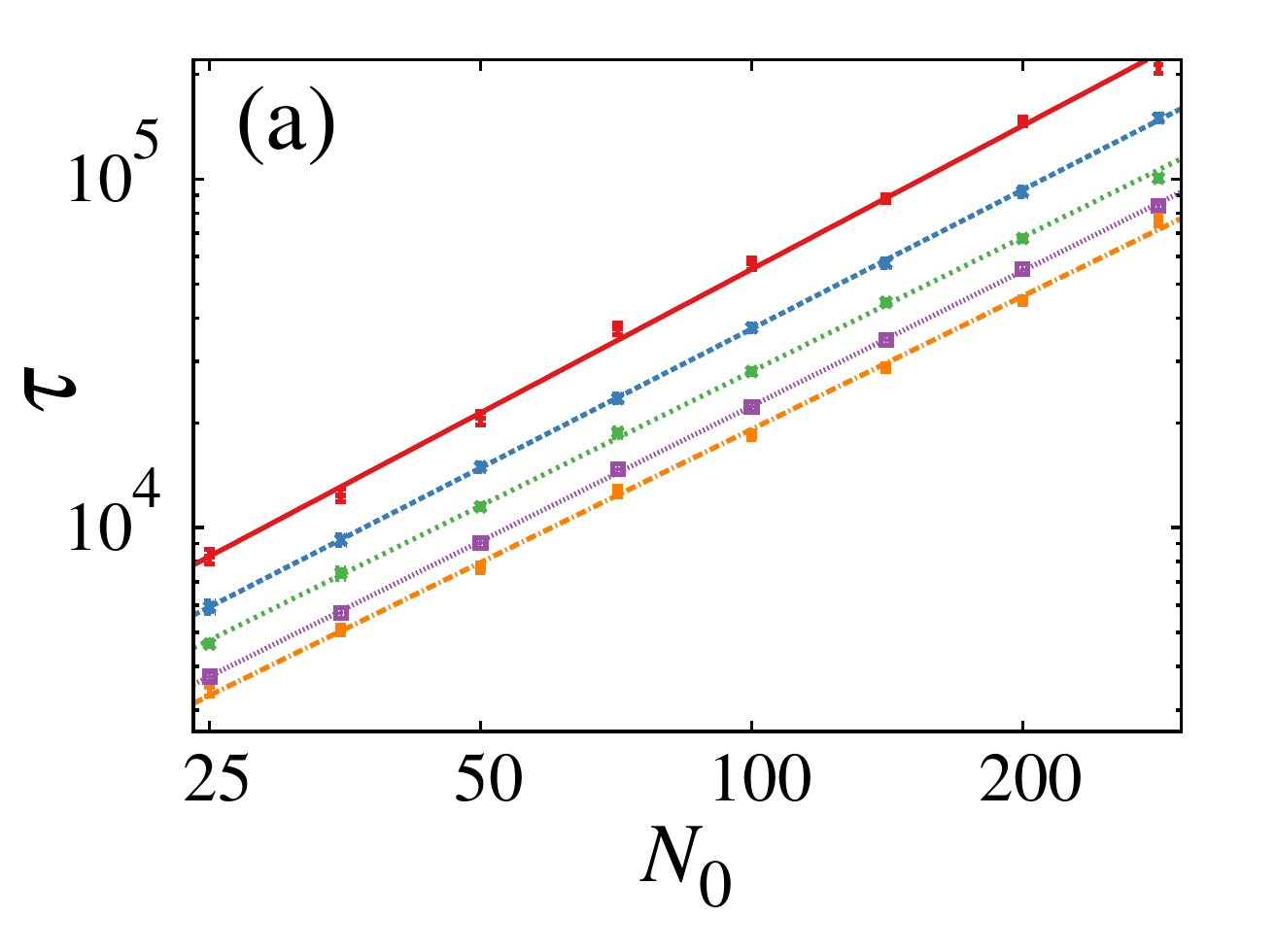}
\includegraphics[width=.33\linewidth]{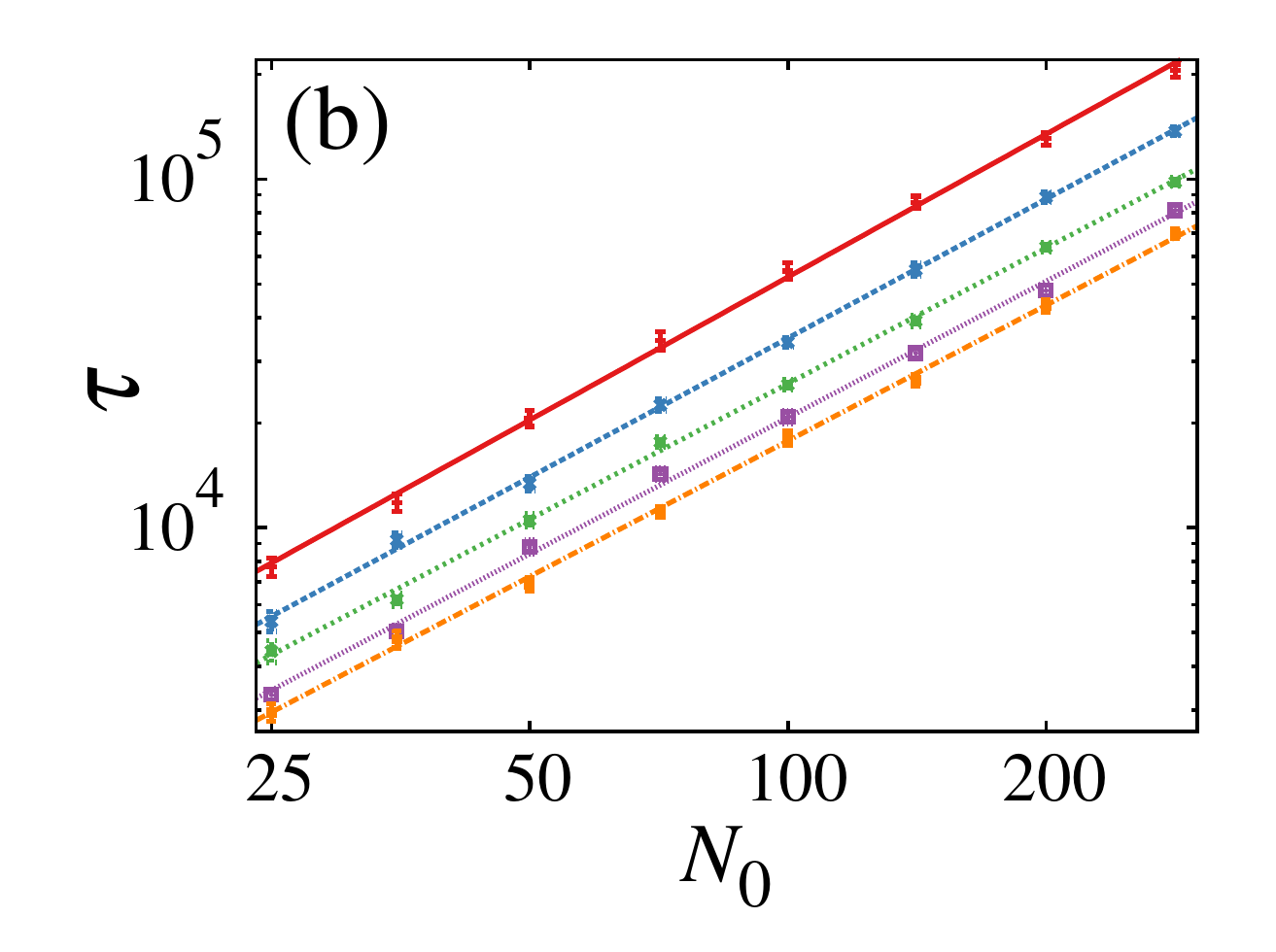}
\includegraphics[width=.33\linewidth]{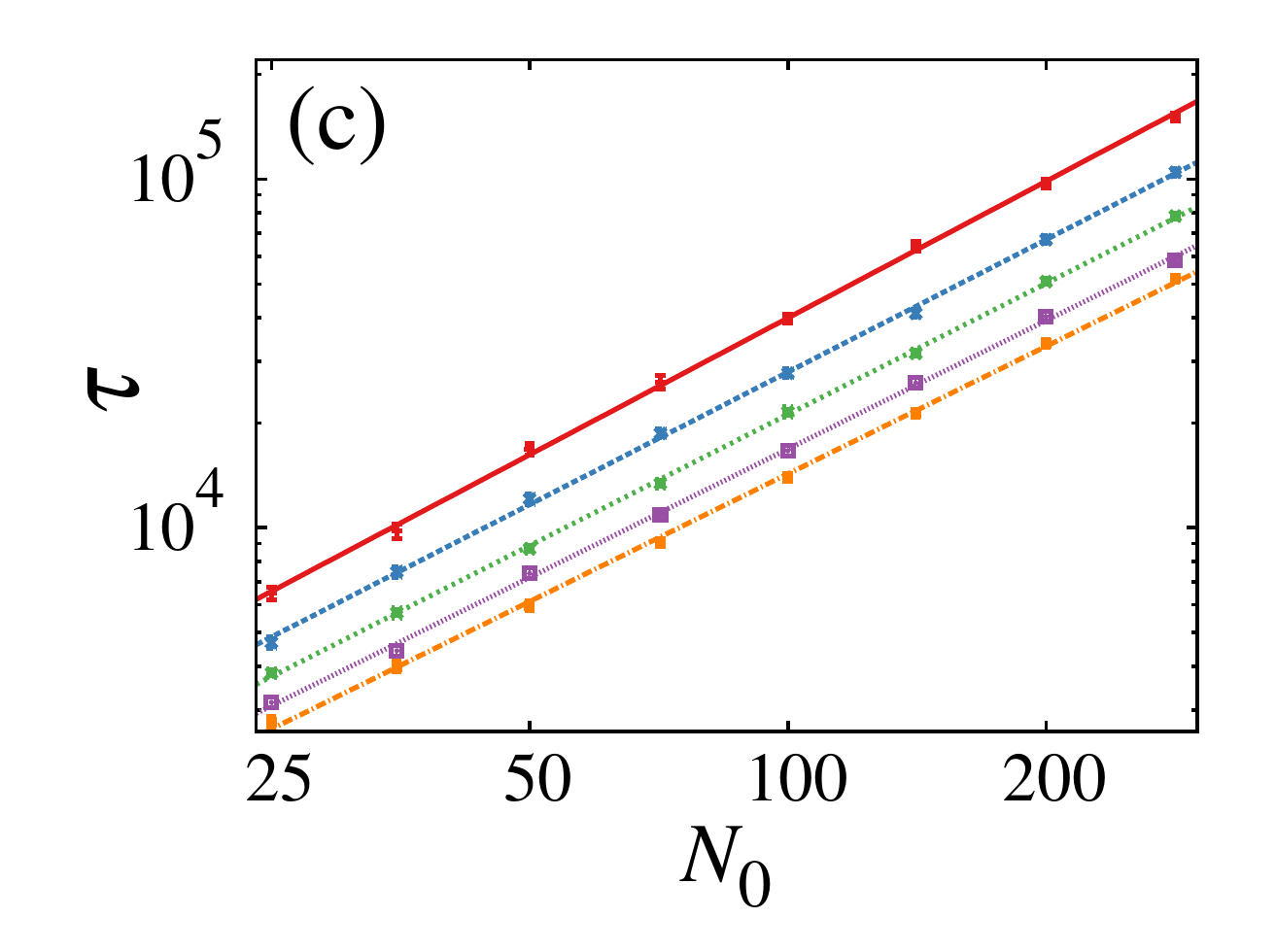}
\includegraphics[width=.33\linewidth]{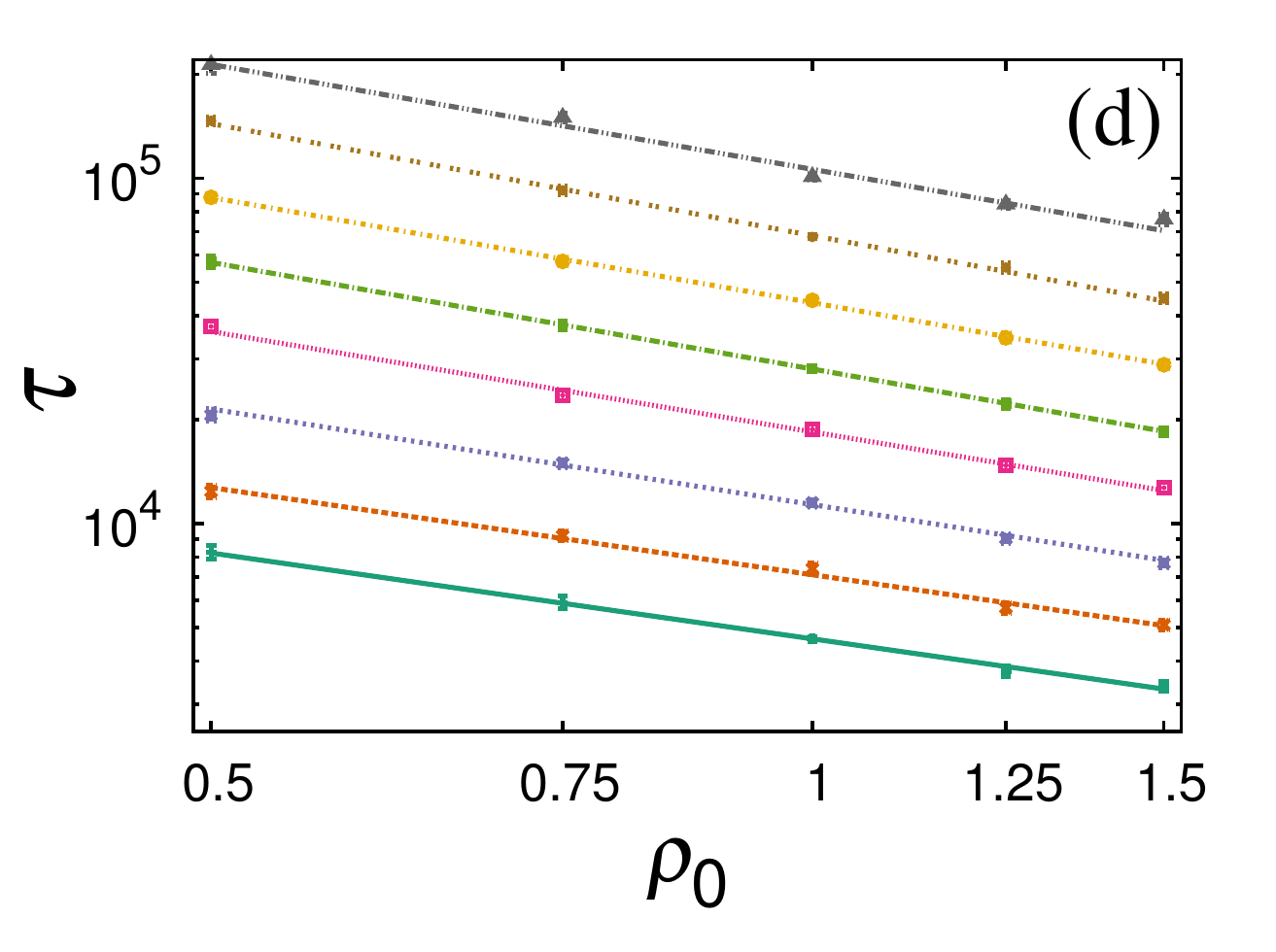}
\includegraphics[width=.33\linewidth]{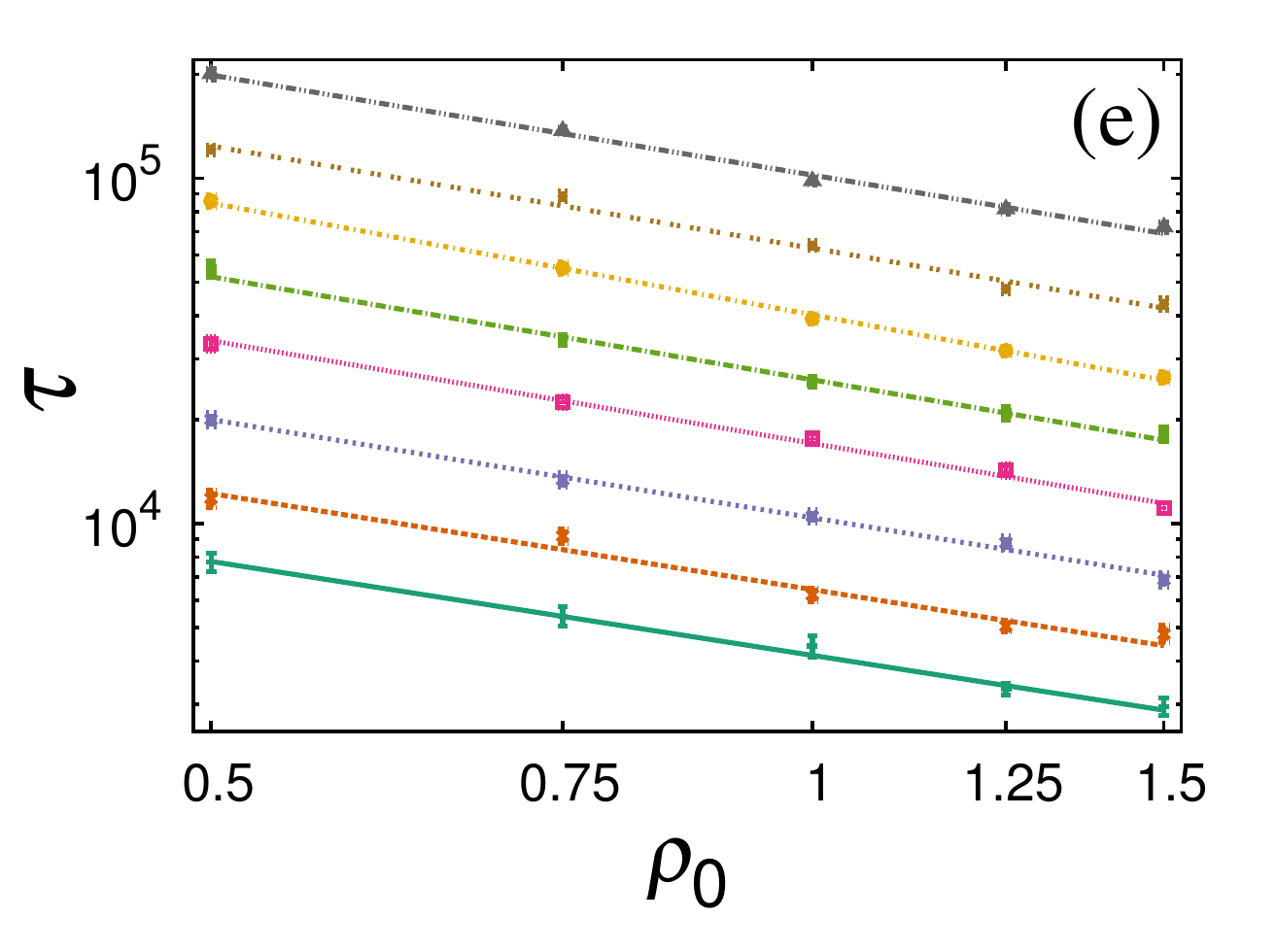}
\includegraphics[width=.33\linewidth]{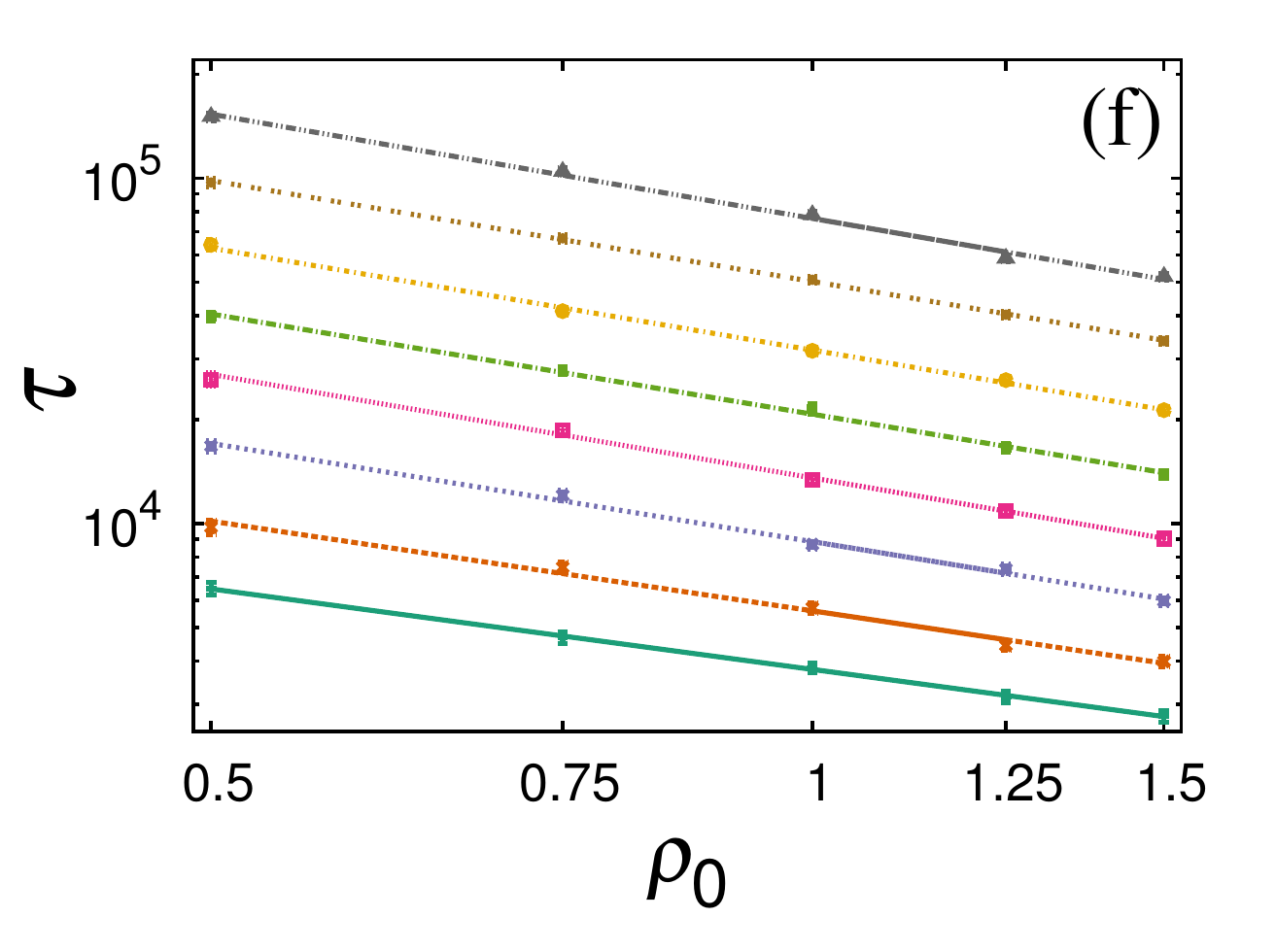}
\caption{(Color online) (a) - (c) Ejection times $\tau$ as a function of polymer length $N_0$ for different initial monomer densities $\rho_0$. Curves from top down: $\rho_0 = 0.5,\ 0.75,\ 1,\ 1.25,$ and $1.5$. (a) LD. (b) SRD without hydrodynamics. (c) SRD with hydrodynamics. Each point is an average over 90-100 runs in (a) and 40-50 runs in (b) and (c). The lines are fits of the form $\tau \sim N_0^{\beta}$. The fitted exponents are tabulated in Table~\ref{tab:exponents}.\\
(d) - (f) Ejection times $\tau$ as a function of initial density $\rho_0$ for polymers of different lengths $N_0$. Curves from top down: $N_0=283,200,141,100,71,50,35,25$. (d) LD. (e) SRD without hydrodynamics. (f) SRD with hydrodynamics. The lines depict the fitting of functions of the form $\tau \sim \rho_0^{-\alpha}$ to the data. The fitted exponents are tabulated in Table~\ref{tab:exponents_rho}. All figures on logarithmic scale.}
\label{fig:tauVsN}
\end{figure*}

\section{Results}\label{sec:res}

In what follows we refer to the SRD method with hydrodynamics as 'with hydrodynamics' or 'HD'. The model where SRD is used but without hydrodynamic interactions is referred to as 'without hydrodynamics' or 'noHD'. The LD method used as a reference does not include hydrodynamic interactions. The presented results are obtained by averaging over typically $50$ runs. For waiting time profiles $500$ ejections were simulated.

\subsection{Ejection time}

Polymer translocation processes are typically characterized by how the translocation time, here also called the ejection time, $\tau$ depends on the  polymer length $N_0$. For the case of translocation through a nanometer-scale pore from one semi-infinite space to another it is established that $\tau \sim N_0^{\beta}$. Ejection time measurements would suggest that such a scaling relation would describe also polymers' ejection from capsids. In our previous study using SRD without hydrodynamics we showed that actually $\beta\rightarrow 1$ as $N_0$ increases~\cite{piili_capsid}. As this contradicts the available theoretical treatments for the capsid ejection starting from moderate monomer densities $\rho_0$~\cite{muthukumar1,cacciuto2,sakaue_polymer_decompression}, a verification using a more established method is called for. We make a close comparison of the polymer ejection models based on SRD and the well established LD. $\tau$ vs $N_0$ for the three different models are shown in Figs.~\ref{fig:tauVsN} (a)-(c). The exponents extracted for the apparent relation $\tau \sim N_0^{\beta}$ are given in Table~\ref{tab:exponents}. LD and SRD without hydrodynamics are seen to give essentially identical scaling. Hydrodynamics is seen to reduce $\beta$ as has been found also for driven polymer translocation~\cite{lehtola_epl}. In both cases hydrodynamic interactions reduce the effective friction the polymer experiences outside the pore. Consequently, the effect of the pore friction, largely caused by the geometry, increases when hydrodynamics is included. Increasing this friction local to the pore with respect to the total friction takes the polymer translocation and ejection toward the linear dependence $\tau \sim N_0$, which explains the reduction of $\beta$ due to hydrodynamics.

Analogously to the case of driven translocation, where translocation time depends on the driving pore force $f_d$ as $\tau \sim f_d^{-\alpha}$, the ejection time decreases with increasing initial density as $\tau \sim \rho_0^{-\alpha}$, see Figs.~\ref{fig:tauVsN} (d)-(f) and Table~\ref{tab:exponents_rho}. For both LD and SRD  $\alpha \to 1$ as $N_0$ increases. Inclusion of hydrodynamics decreases $\alpha$, again in analogy with driven translocation~\cite{lehtola_epl}. This is accounted for by the hydrodynamic interactions decreasing the effective length of the polymer due to increased correlation length along the polymer.

In summary, mere ejection time measurements would seem to confirm previous results on $\tau$ scaling with polymer length $N_0$. Also, the dependence of $\tau$ on the initial monomer density $\rho_0$ would seem to corroborate the scaling behavior. Results using SRD and LD are essentially identical. The theoretical arguments have typically been corroborated by ejection time measurements alone. However, inspection of the measured waiting time profiles changes the conclusions completely.

\begin{table}
\caption{The exponents $\beta$ of the fits to the apparent relation $\tau \sim N_0^\beta$ plotted in Fig.~\ref{fig:tauVsN}~(a),(b), and (c) for different initial monomer densities $\rho_0$. The errors of the fits are of the order 0.04.}
\begin{tabular}{|c|c|c|c|c}
\hline
$\rho_0$ & HD & noHD & Langevin \\\hline
0.50 & 1.30 & 1.36 & 1.37 \\
0.75 & 1.26 & 1.33 & 1.33 \\
1.00 & 1.25 & 1.29 & 1.28 \\
1.25 & 1.22 & 1.30 & 1.29 \\
1.50 & 1.22 & 1.29 & 1.27 \\\hline
\end{tabular}
\label{tab:exponents}
\end{table}

\begin{table}
\caption{The exponents $\alpha$ of the fits $\tau \sim \rho_0^{-\alpha}$ plotted in Figs.~\ref{fig:tauVsN}~(d),(e), and (f) for different polymer lengths $N_0$. The errors of the fits are of the order 0.1.}
\begin{tabular}{|c|c|c|c|c}
\hline
$N_0$ & HD & noHD & Langevin \\\hline
25 & 0.77 & 0.89 & 0.83 \\
35 & 0.85 & 0.89 & 0.82 \\
50 & 0.94 & 0.96 & 0.90 \\
71 & 0.98 & 1.00 & 0.97 \\
100 & 0.96 & 1.00 & 1.03 \\
141 & 0.99 & 1.08 & 1.01 \\
200 & 0.96 & 1.05 & 1.07 \\
283 & 1.00 & 0.99 & 0.98 \\\hline
\end{tabular}
\label{tab:exponents_rho}
\end{table}

\begin{figure*}
\includegraphics[width=.33\linewidth]{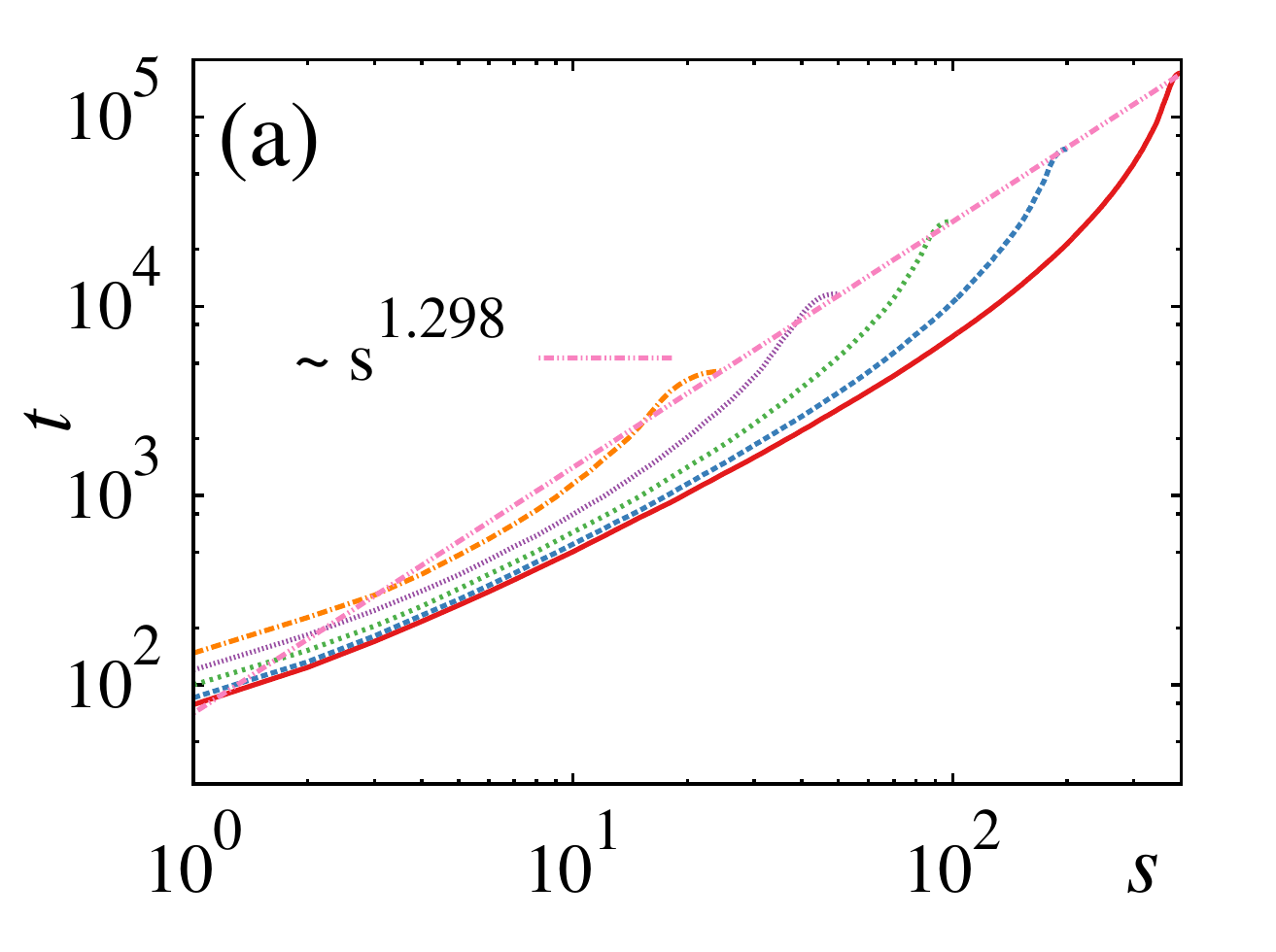}
\includegraphics[width=.33\linewidth]{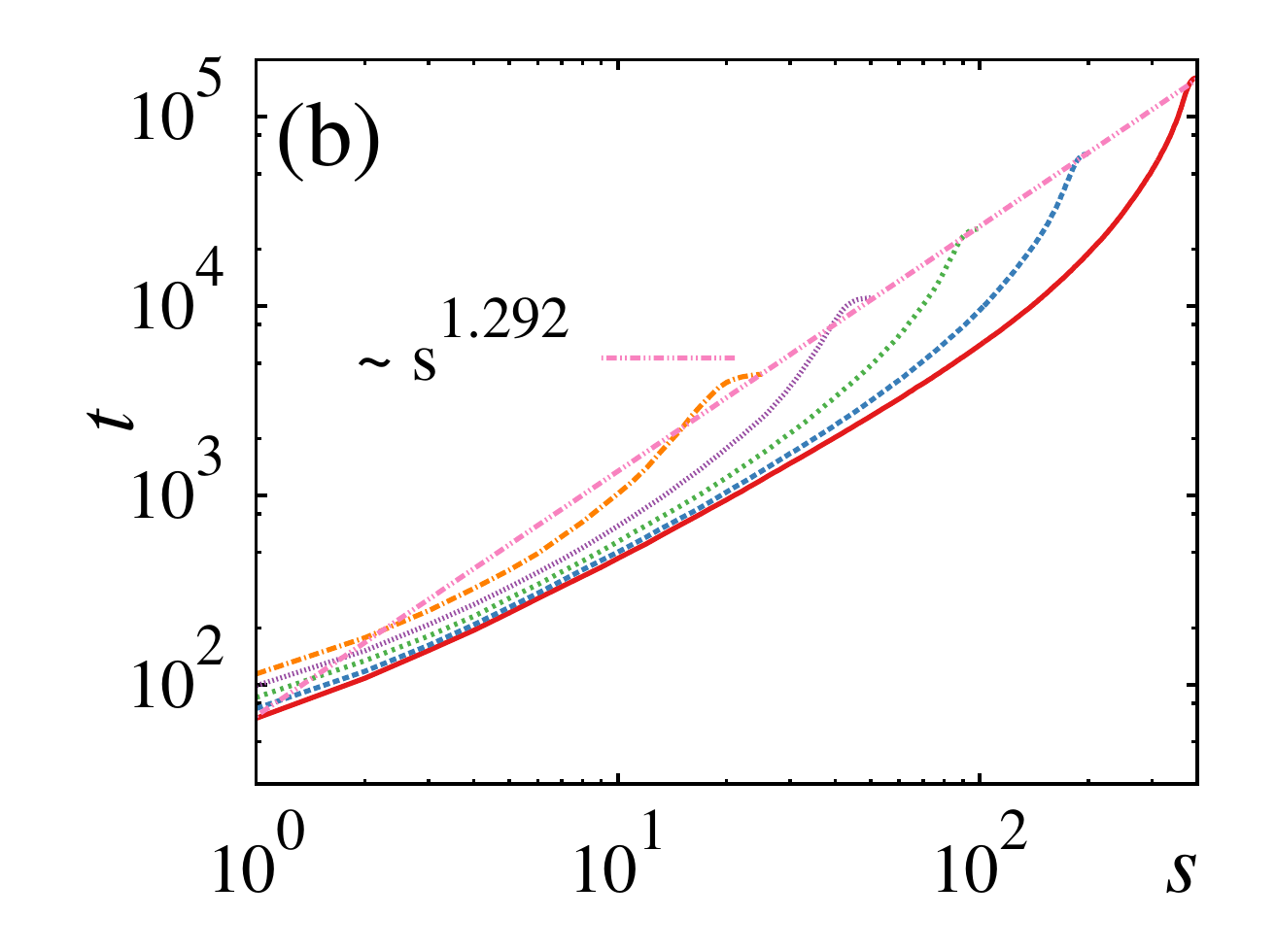}
\includegraphics[width=.33\linewidth]{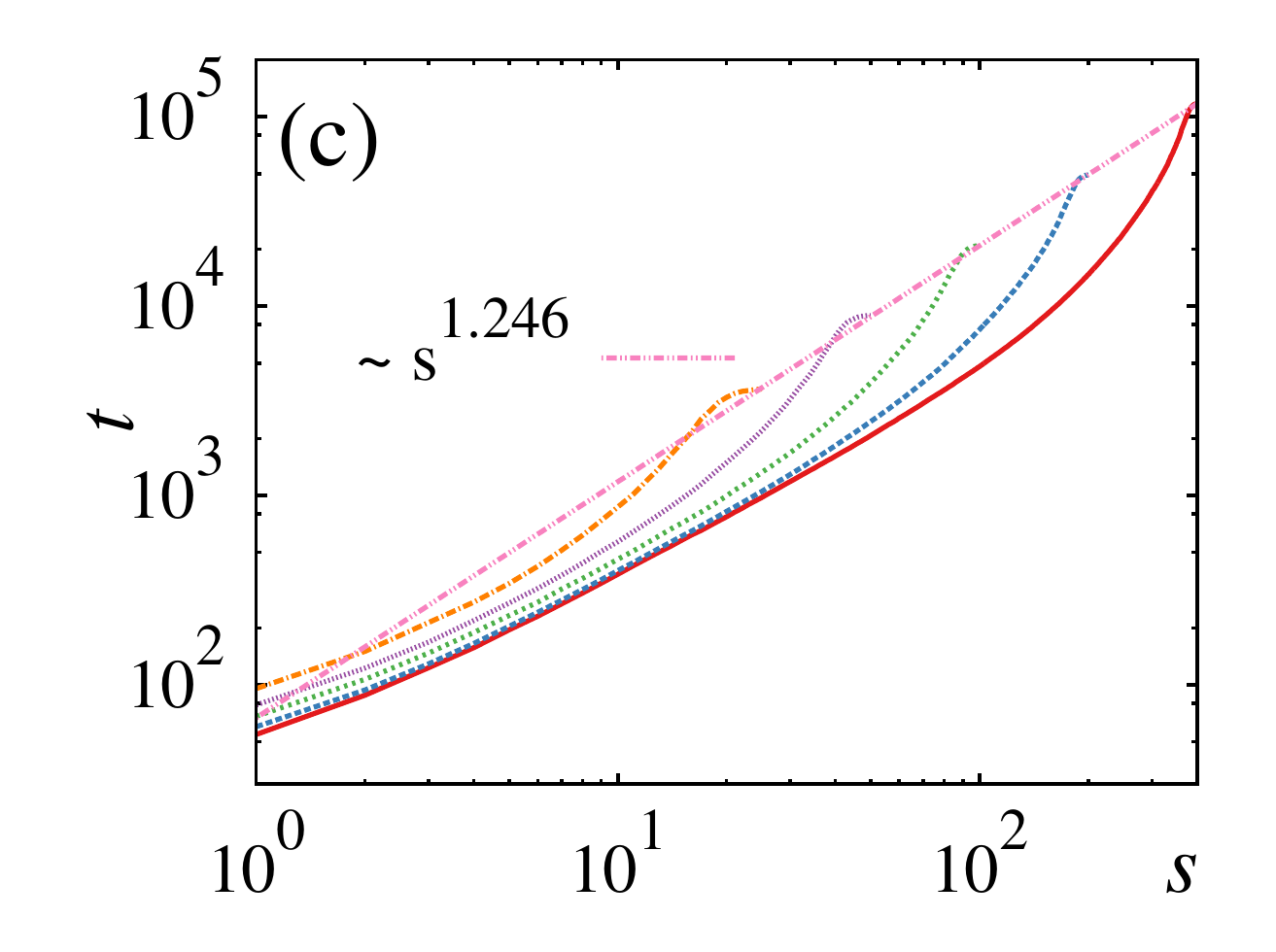}
\includegraphics[width=.33\linewidth]{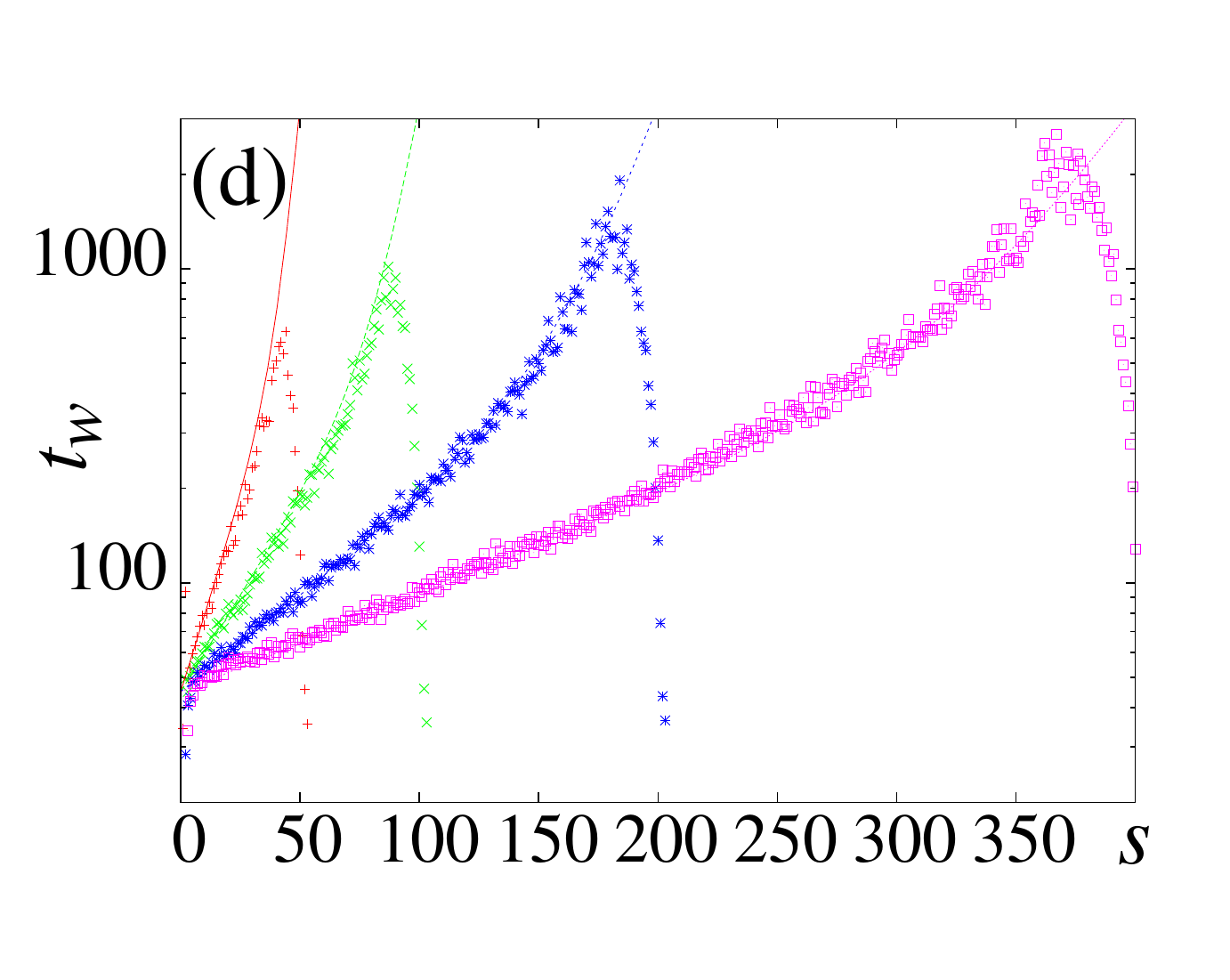}
\includegraphics[width=.33\linewidth]{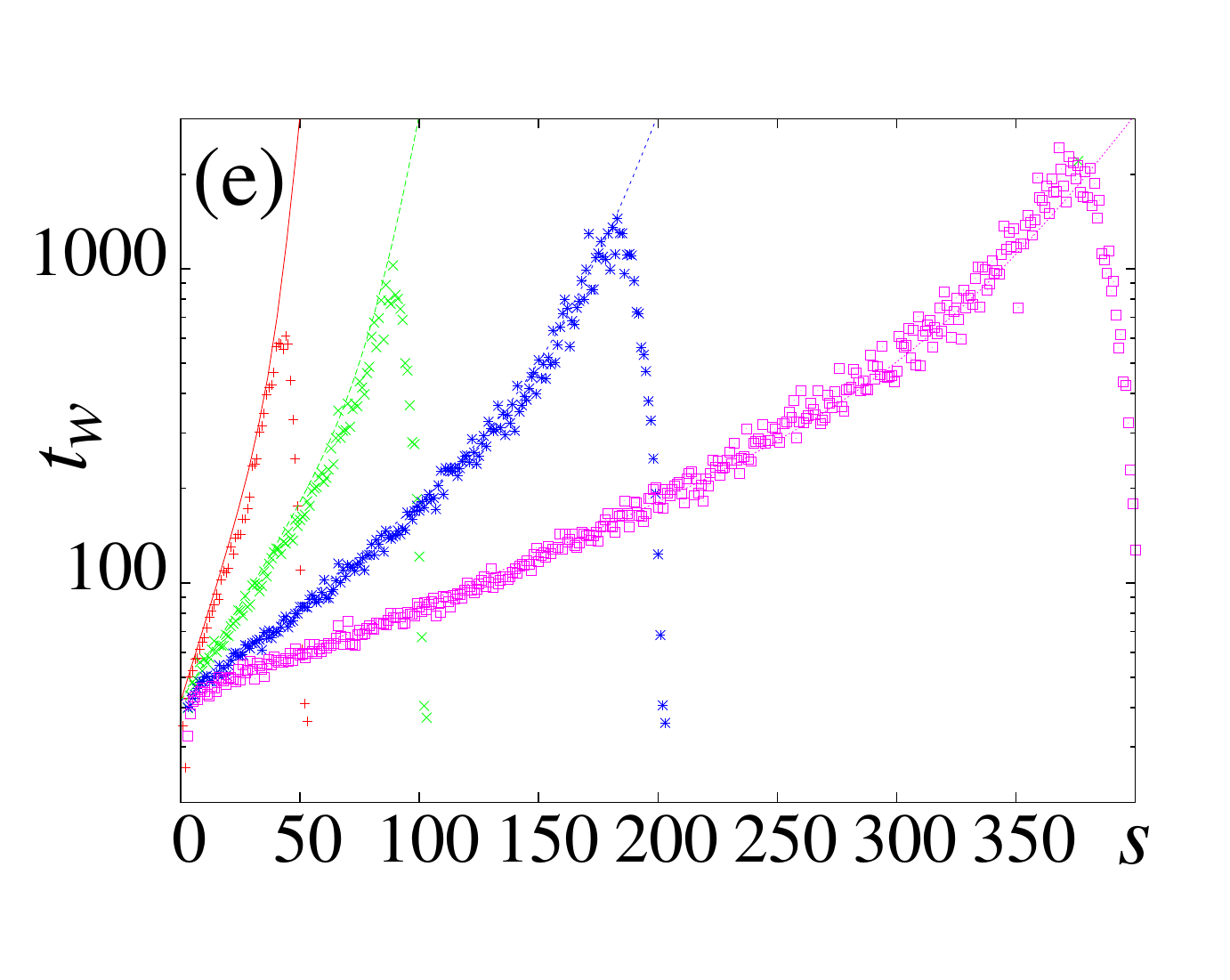}
\includegraphics[width=.33\linewidth]{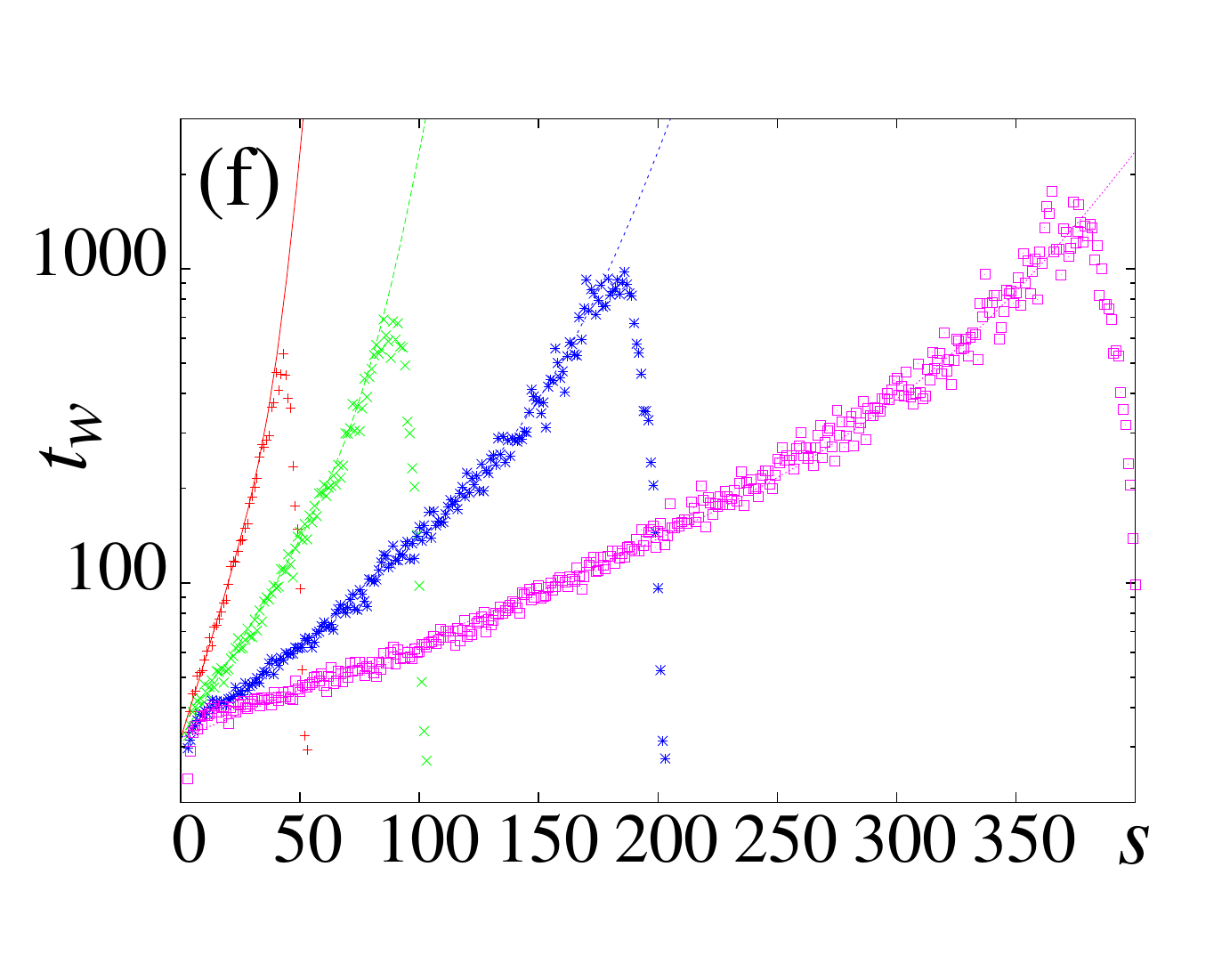}
\caption{(Color online) Cumulative and differential waiting times. From left to right: LD, SRD without hydrodynamics, and SRD with hydrodynamics. Initial monomer density $\rho_0=1.0$. $N_0 = 50,\ 100,\ 200,$ and $400$. In (a)-(c) also $N_0=25$ is included.\\
(a)-(c) Cumulative waiting times $t$ as a function of the reaction coordinate $s$ on a logarithmic scale. The dashed lines show the scaling of the endpoints.\\
(d)-(f) Waiting times $t_w$ and fits of the form $t_w = A\left[\exp{\left((2.8/N_0)s\right)}+\exp{\left((10.8/N_0)(s-0.625N_0)\right)}\right]$, where (d) $A = 45$, (e) $A = 42$, and (f) $A = 32$. Semilogarithmic scale.}
\label{fig:waitingTimesAll}
\end{figure*}
\begin{figure}[b]
\includegraphics[width=0.49\linewidth]{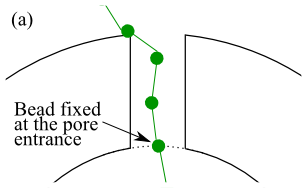}
\includegraphics[width=0.49\linewidth]{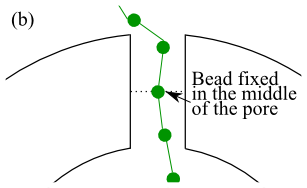}
\caption{(Color online) The two ways used to measure the force at the pore. The force required to hold the monomer either (a) at the entrance or (b) in the middle of the pore is measured.}
\label{fig:fdiagram}
\end{figure}
\begin{figure*}[htb!]
\includegraphics[width=.33\linewidth]{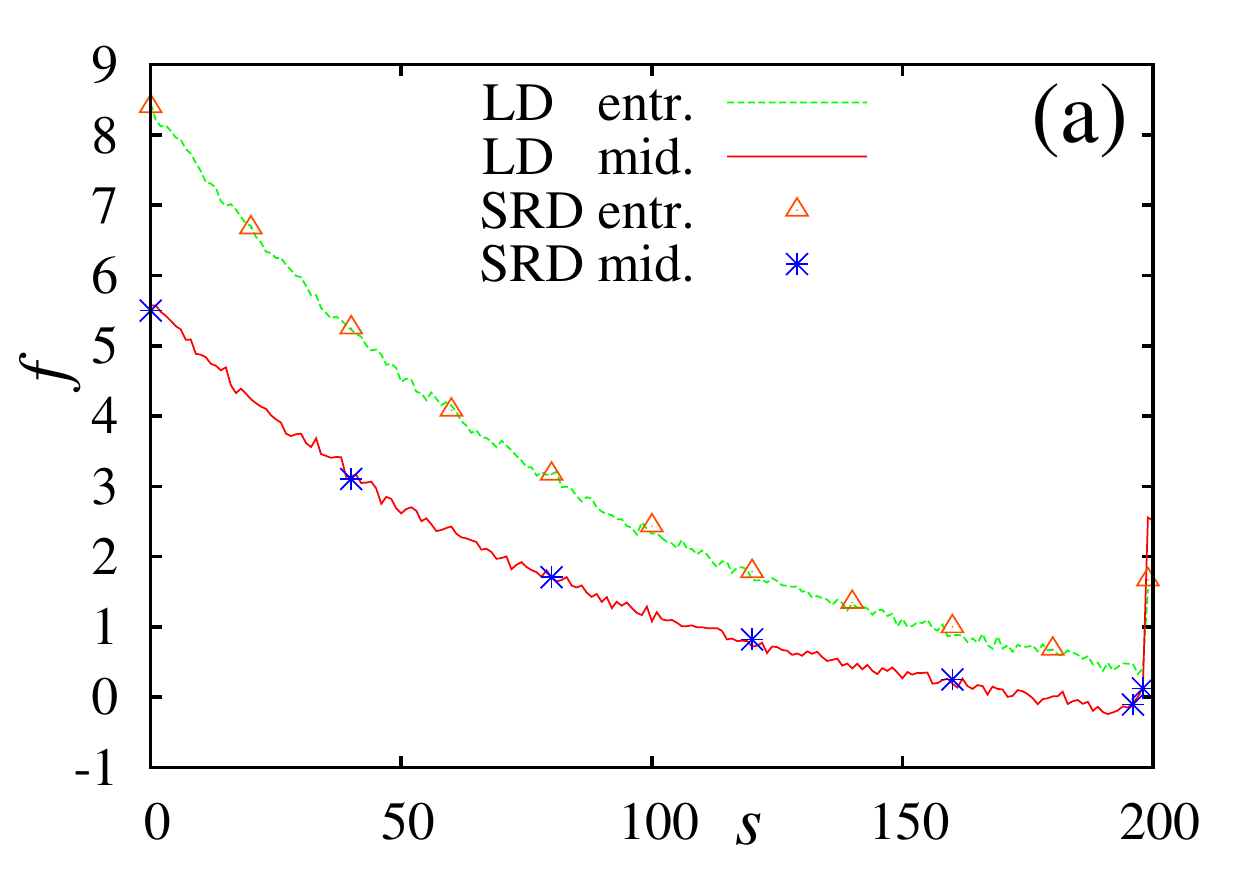}
\includegraphics[width=.33\linewidth]{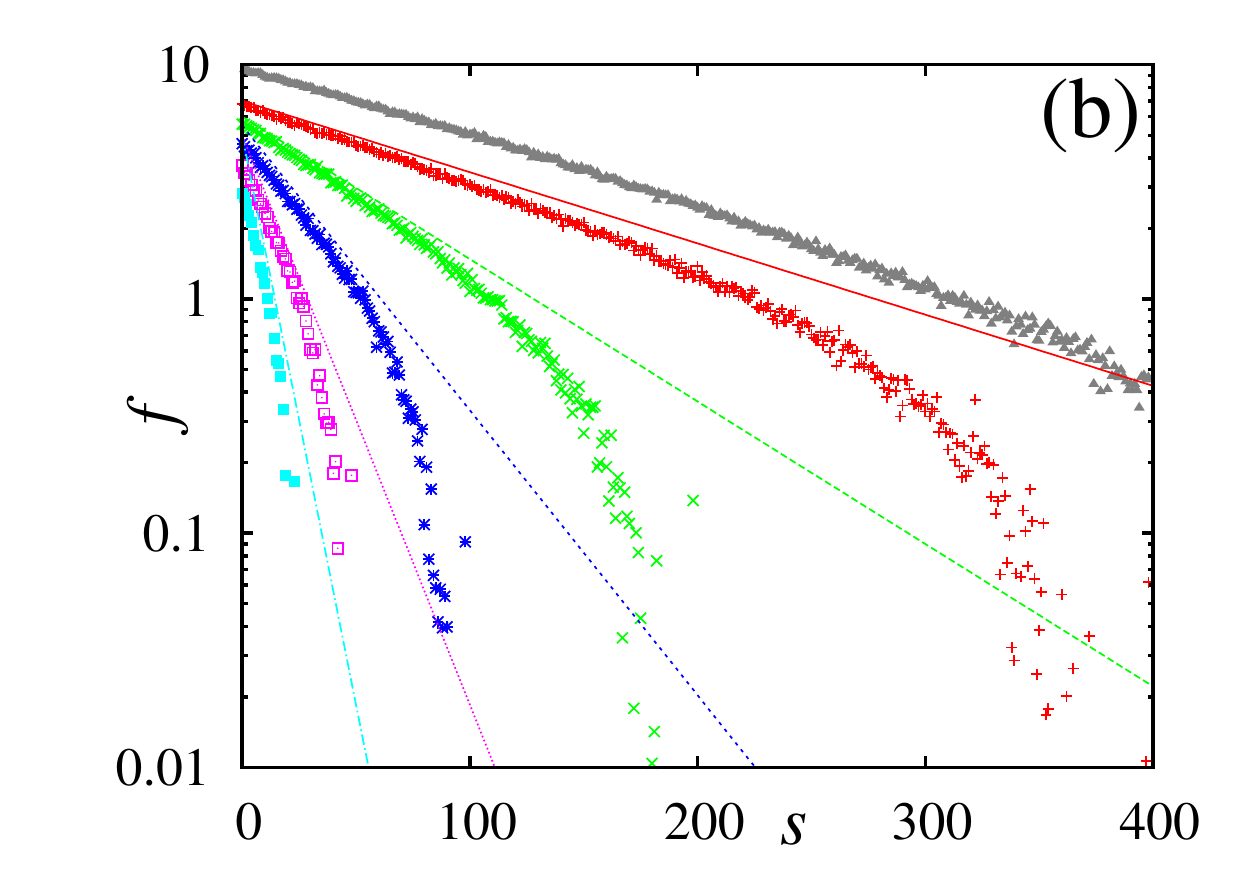}
\includegraphics[width=.33\linewidth]{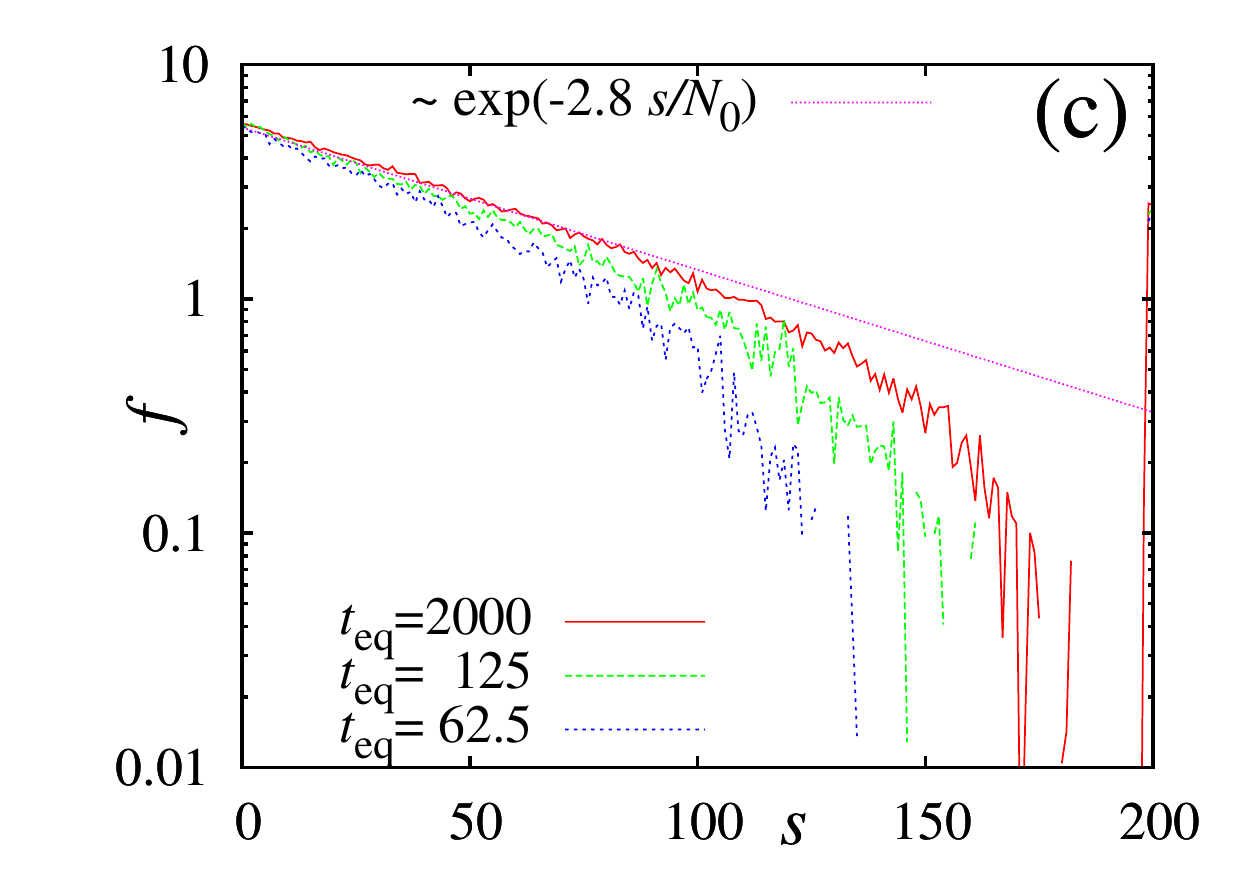}
\caption{(Color online) (a) Force $f$ measured in the SRD  and LD models at the {\it cis} side entrance and the symmetrical midpoint of the pore as a function of the translocation coordinate $s$. $N_0=200$ and $\rho_0=1.0$. (b) Force $f(s)$ measured in the middle of the pore in the LD model for $N_0 = 25,\ 50,\ 100,\ 200,$ and $400$. The topmost curve shows $f(s)$ measured at the pore entrance for $N = 400$. The lines plot $f \propto \exp((-2.8/N_0)s)$ for the different $N_0$ (see text). (c) Force $f(s)$ when the bead $s$ at the pore opening was held at its place for time $t_{\rm eq}$ after which $f(s)$ was measured for time $t_{\rm m} = t_{\rm eq}$. The curves are from top to bottom $t_{\rm eq} = 2000,\ 125,$ and $62.5$.}
\label{fig:force}
\end{figure*}
\subsection{Waiting times}

In this section we extract the waiting time profiles $t_w(s)$ for the different models. We obtain a more precise form for $t_w(s)$ than in our previous study~\cite{piili_capsid} and find that it is universal for all models.

$t_w(s)$ is defined as the time it takes for the bead $s$ to eject the capsid after the ejection of the previous bead $s-1$.
\begin{align}
t_w(s) = t(s) - t(s-1),
\end{align}
where $t(s)$ is the time when the bead $s$ exits the capsid for the last time, that is, the cumulative waiting time. $t_w(s)$ plotted as a function of $s \in [1,N_0-1]$ is the waiting time profile. For the process to be genuinely scale-invariant with respect to the polymer length, both $t(s)$ and $t_w(s)$ should scale with $s$. Figures~\ref{fig:waitingTimesAll}~(a)-(c) show $t(s)$ obtained for the three models. For all the models the endpoints $t(N_0-1)$ scale with $N_0$ in accordance with the apparent scaling relation $\tau \sim N_0^\beta$. However, $t(s)$ do not scale with $s$.

Figures~\ref{fig:waitingTimesAll}~(d)-(f) show the waiting time profiles $t_w(s)$ for the three different models. $t_w(s)$ is seen to be of the common form

\begin{align}\label{wt}
t_w(s) = A\left[\exp{\left(\frac{2.8}{N_0}s\right)} + \exp{\left(\frac{10.8}{N_0}\left(s-0.625N_0\right)\right)}\right],
\end{align}
where $A = 45,\ 42,$ and $32$ for LD, SRD without hydrodynamics, and SRD with hydrodynamics, respectively. Hence, for all the models polymer ejection slows down exponentially with the length of the ejected segment $s$. At a definite stage when approximately $63 \%$ of the polymer has been ejected the ejection slows down more strongly with $s$. Presumably, the transition corresponds to $s = s_0$ when the monomer density inside the capsid is so low that internal pressure no longer exerts force on the ejecting polymer, see Section~\ref{sec:force}. The exponential form $t_w \sim \exp(C s)$ can only lead to linear dependence of the ejection time with $N_0$ for long polymers~\cite{piili_capsid}. This is true also for the sum of two exponential functions as in Eq.~(\ref{wt}). Consequently, for sufficiently long polymers there is a scaling function $h$ such that $t(s) = N_0 h(s/N_0)$. In the present case 
\begin{align}
h\left(\frac{s}{N_0}\right) = & A \left\lbrace\frac{1}{2.8} \exp\left(\frac{2.8 s}{N_0}\right)\right.\nonumber\\
&\left. + \frac{1}{10.8} \exp\left[\frac{10.8}{N_0}\left(s-0.625 N_0\right)\right]-\text{const.}\right\rbrace.
\end{align}

It is seen that hydrodynamics only reduces the magnitude of waiting time profile $t_w(s)$ without changing its form. In other words, the waiting times are related via $t_w^{\rm HD}(s) = (\tau_{\rm HD}/\tau_{\rm noHD}) t_w^{\rm noHD}(s)$, for the $N_0$ and $\rho_0$ selected. This is reminiscent of the driven polymer translocation where hydrodynamics speeds up translocation and scales down the length of the tensed segment $l$ on the {\it cis} side without changing the way the tension spreads on the polymer chain, that is, the form of $l(s)$~\cite{jaakko}. Furthermore, $t_w(s)$ obtained by using LD aligns almost perfectly to that given by SRD without hydrodynamics within the precision of the mapping of the two models via the friction parameter, see Section~\ref{sec:langevinMapping}. Polymer ejects slightly faster in SRD without hydrodynamics than in LD simulations even though LD has a slightly smaller friction parameter $\xi$ in free solvent, as measured in Section~\ref{sec:langevinMapping}. This would indicate that for high monomer concentrations SRD has enhanced correlations between polymer beads residing in the same cell which would decrease the effective friction.

\subsection{Force measured at the pore}\label{sec:force}
Thanks to the implementation of the capsid geometry using constructive solid geometry no explicit forces at the pore are imposed. Hence, the ejection force results as far as possible from the pressure of the polymer segment confined inside the capsid. There remains contribution from local effects, such as the sharp edges of the pore restricting polymer movement at both openings, but these can be considered relevant also to real pores.

We characterize the dynamic force at the pore by measuring the force $f$ for fully and partly equilibrated polymer conformations for different reaction coordinate $s$. In SRD polymers of different lengths $N_0$ are packed to a random conformation inside a capsid of inner volume $V = 4/3\pi R_0^3 = N_0/\rho_0$ until the bead $s$ is at the pore entrance. Here $\rho_0$ is the initial monomer density of a corresponding capsid ejection.

In our previous study~\cite{piili_capsid} the bead $s$ was attached to a point in the pore entrance via a FENE potential and the average force needed to keep the polymer in place was measured, see Fig.~\ref{fig:fdiagram}~(a). In response to the results from more precise measurements using LD, reported in what follows, we change here the measurement point to the middle of the pore, see Fig.~\ref{fig:fdiagram}~(b). In SRD, after attaching the bead at either the entrance or the middle of the pore we  wait for a time $t_{\rm eq} = 2.2 \cdot 10^4$ before measuring the force over $t_m = 2 \cdot 10^4$ time steps, one measurement per step, and averaging over them. For a few $s$ we checked that setting the equilibration time $t_{\rm eq} = 5 \cdot 10^4$ did not change the measured average $f$. Hence, during the measurement the polymer is at or very close to an equilibrium conformation. 

The measured force for equilibrated polymer conformations are not equal to the dynamic force during ejection. In order to be able to measure $f$ for conformations that are not fully equilibrated we do the measurement in a slightly more complicated way in the LD model. Here the polymer is initially packed inside the capsid and then freed for a single bead to eject. The appropriate bead is then pulled either to the pore opening or the middle of the pore and held fixed for a time $t_{\rm eq}$ after which the harmonic force that is needed to keep the bead fixed is measured for a time $t_m = t_{\rm eq}$. After this the polymer is again freed for a single bead to eject. This way force is measured for all $s$.

First we verify that the same equilibrium $f$ is obtained using SRD and LD. $f(s)$ for fully equilibrated conformations measured in the SRD model and in the LD model using $t_{\rm eq} = t_m = 2000$ are shown in Fig.~\ref{fig:force}~(a). $f(s)$ for both models are seen to be identical. Force measured at the pore entrance $f_{\rm ent}(s)$ is seen to be larger than force measured in the middle of the pore $f_{\rm mid}(s)$; it does not decay to zero even at the end of the ejection. This is caused by the reduction of the degrees of freedom at the pore entrance due to the measured bead being held stationary there. This creates a bias toward the exit of the pore. The surprisingly strong bias created by an asymmetry in the pore was noticed already in~\cite{riku_dynamics_of_ejection}.

Eliminating the bias due to asymmetry imposed by the measurement affects the dependence of $f$ on $s$. Figure~\ref{fig:force}~(b) shows $f_{\rm mid}(s)$ for polymers of different lengths in the LD model together with $f_{\rm ent}(s)$ for $N_0 = 400$. Measurements are done for $t_{\rm eq} = t_m = 2000$, so these $f$ are for conformations that are close to equilibrium. For the initially large monomer density $\rho$, that is, for relatively small $s$, all $f$ show close to exponential decay with $s$. This decay rate is very close to the rate of the exponential increase of the waiting time $t_w$ with $s$ for $s \le 0.625 N_0$, see the first exponential term in Eq.~(\ref{wt}). The lines in Fig.~\ref{fig:force}~(b) show the function $f \propto \exp(-(2.8/N_0)s)$ for the different $N_0$. For $s \le 0.625 N_0$ the ejection dynamics is thus seen to mainly result from the exponential decay of the pressure inside the capsid. Finally, as the pressure decreases more abruptly, there is a crossover to a stronger exponential increase of $t_w$, see Eq.~(\ref{wt}).

The exponential decay of $f$ coinciding with the exponential increase of $t_w$ for $s \le 0.625 N_0$, that is, for the part of the process where the pressure resulting from packed monomers drives the ejecting polymer, is in keeping with our previous findings. For these densities and polymer lengths the number of beads per blob is very low, so the blob picture used in many theoretical approaches is not relevant but monomers interact individually. The exponential dependence of the resulting potential inside the capsid and hence the driving force has an exponential dependence on monomer density and so $s$. This can be obtained from the Flory-Huggins theory mixing free energy in the limit of high density~\cite{piili_capsid}. For large monomer densities and small $s$ we also find that $t_w(s) \sim 1/f(s)$. Observe that the small deviation from inverse proportionality to a power law relation $t_w(s) \sim 1/f(s)^\gamma$, where $\gamma\approx 0.95$, leads to the slightly deviating exponent as a form of $f(s) \sim \exp{\left(-\frac{2.8}{\gamma N_0} s\right)}$. This explains the slightly different exponent in the force curves in Fig.~\ref{fig:force}~(b).

The waiting time shows a stronger exponential for lower monomer densities and large $s$, see Eq.~(\ref{wt}). Fig.~\ref{fig:force}~(c) shows force measured at the pore midpoint using LD and allowing for the polymers to equilibrate for different times $t_{\rm eq} = t_{\rm m}$, as explained above. It is seen that the further the polymer conformation is from the equilibrium, the more abruptly the measured $f$ falls off with increasing $s$ similarly to the corresponding stronger increase of $t_w$ with $s$ for $s > 0.625 N_0$. There are two potential reasons for this: First, on the {\it cis} side at the final stage tension may propagate in the remaining polymer segment, which increases friction and diminishes force measured at the pore. Second, on the {\it trans} side monomers may crowd thus possibly impeding the ejection of the polymer. We have shown that crowding plays no role in driven polymer translocation~\cite{pauli_translocation}. The force in the final ejection stage is much smaller than in the driven translocation, so here the effect cannot be ruled out offhand. The relevant non-equilibrium mechanism is determined in Section~\ref{modmodels}.
\begin{figure}
\includegraphics[width=0.8\linewidth]{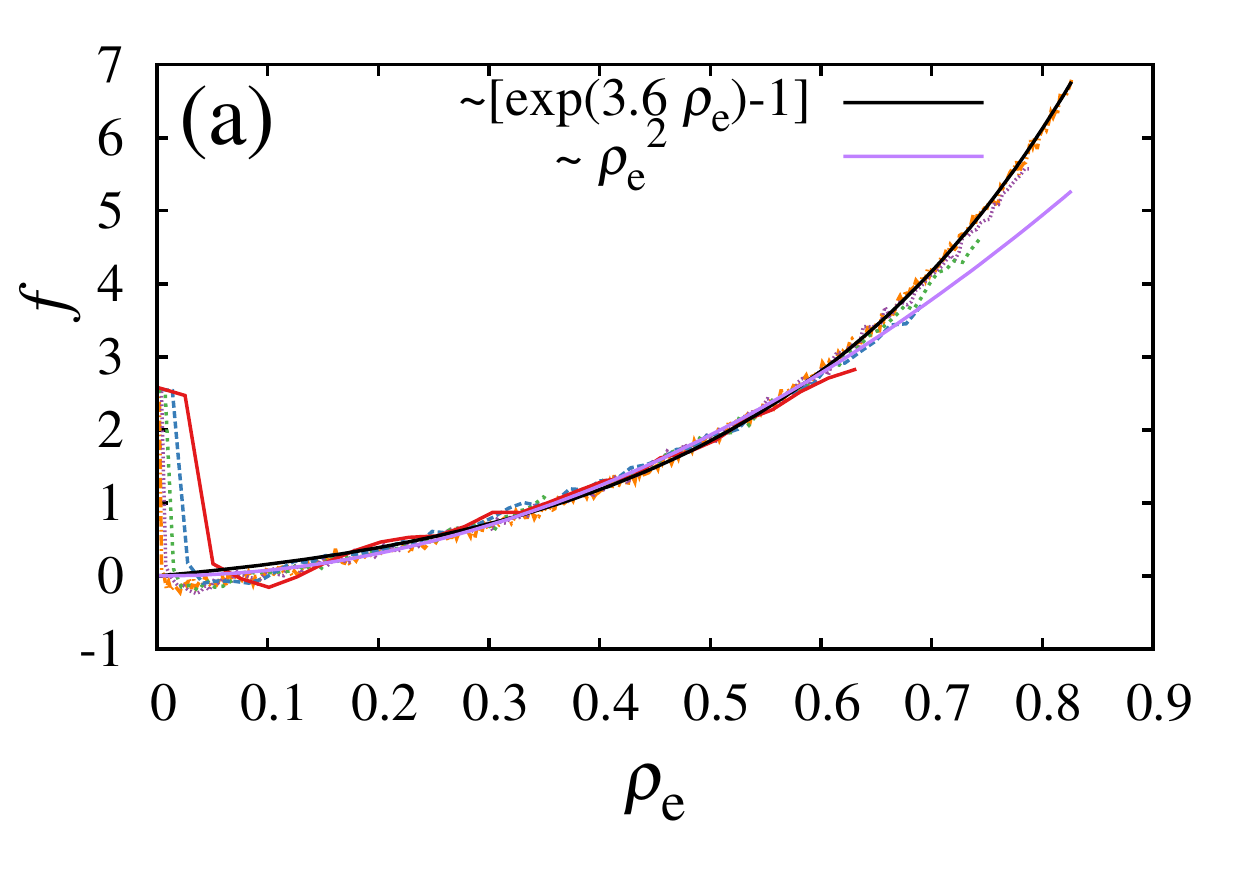}
\includegraphics[width=0.8\linewidth]{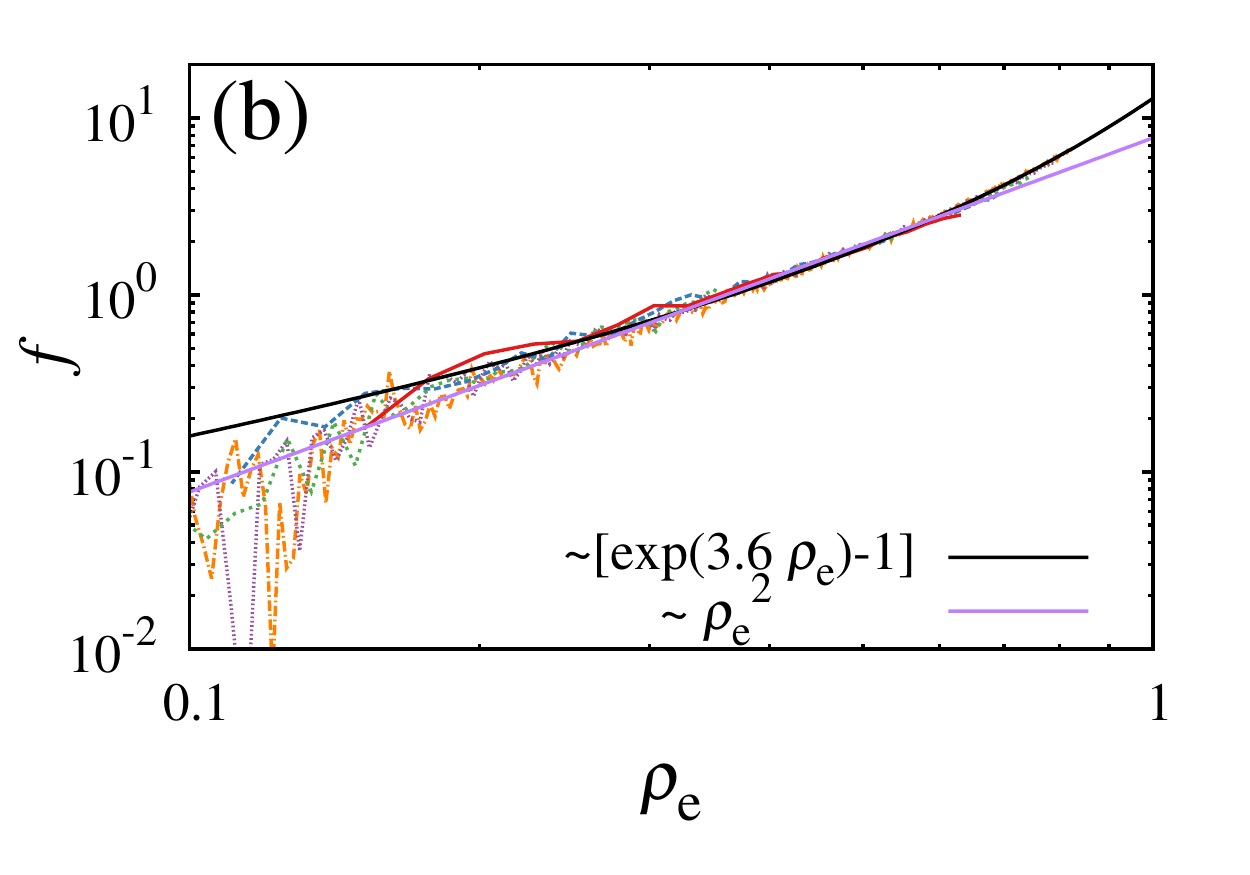}
\caption{(Color online) Force $f$ as a function of the effective monomer density $\rho_e$ inside the capsid for all $s$ for polymer lengths $N_0=25,\ 50,\ 100,\ 200,$ and $400$ with $\rho_0=1.0$ (a) natural scale (b) logarithmic scale. Different scales and fits are shown to emphasize the exponential dependence for higher effective densities and the potential scaling relation for lower densities.}
\label{fig:rho}
\end{figure}

The greater statistics due our LD model enables us to extract the dependence of the equilibrium force measured in the middle of the pore on the monomer density inside the capsid more precisely than using SRD. We find that the apparent dependence on the polymer length arises from the inevitable overlap of the repulsive monomer potentials with the capsid wall. This overlap is proportionally larger for small capsids and short polymers. We introduce the volume correction parameter $\epsilon$ to take this into account. The effective monomer density then becomes

\begin{align}\label{equ:rhoeff}
 \rho_{e} = \frac{N}{V_{ e}} = \frac{N}{\frac43 \pi (R_0 + \epsilon)^3},
\end{align}
where $N$ is the number of monomers inside the capsid. The data for the measured $f$ vs $\rho_e$ for different $N_0$ falls onto the same curve when $\epsilon = 0.3$.

Figure~\ref{fig:rho} shows the measured pore force in equilibrium as a function of the effective density inside the capsid $\rho_e$ in natural and logarithmic scales. For large and intermediate monomer densities the force grows exponentially with $\rho_e$ in keeping with the exponential decay of $f$ with $s$. However, for sufficiently small monomer densities the exponential relation does not hold precisely. This is due to the measured force decaying to zero and even going slightly negative for extremely small monomer densities.

In Fig.~\ref{fig:rho} we have fitted to the data the function $f(s) = C\left(\exp{\left(B\rho_e\right)} - 1\right)$ which, unlike a pure exponential, has the property that it decays to zero when the monomer density is exactly zero. This function was chosen because we believe that the negative force is an artefact caused by the local pore geometry and the selected position for the force measurement. In measurements at the \textit{cis} entrance the force is always positive while in the mid-pore measurement the minimum force is negative. This implies that there must be a measurement position where the minimum force is exactly zero. The relatively small offset in the force might seem like a minor detail, but as an ejecting polymer spends the majority of its time in the small force regime, the form of the force has a considerable effect on ejection time.

The form of the pore force leaves some room for speculation. For instance, it is also possible to describe the force in the small $\rho_e$ regime as a power law $f\sim \rho_e^2$ due to the shifted exponential and power law resembling each other in such a short range. Cacciuto and Luijten~\cite{cacciuto_free_energy} measured the scaling of the excess free energy with the number of monomers in the capsid as $\Delta F \sim N^{2.97}$ for $0.2 \le \phi < 0.35$, where $\phi$ is the volume fraction. This would give $f = - \partial \Delta F / \partial N \sim N^{1.97}$, which is approximately the scaling obtained here. In Fig.~\ref{fig:rho}~(b) we have plotted the scaling of this form on logarithmic scale alongside with force measurements. However, the scaling regime obtained here is for lower $\rho$ and cannot be of the same origin as assumed in~\cite{cacciuto_free_energy}, namely the screening that steps in at higher $\rho$.

The main observation here is that in our model we do find a narrow interval at low $\rho$, where $f$ for equilibrium conformations may scale with $\rho$. However, the scaling exponent is not of the same magnitude that could be derived using the blob-scaling arguments. In addition, during the ejection the conformations are out of equilibrium.

\subsection{First passage times derived from the pore force}
\begin{figure}
\centering
\includegraphics[width=0.9\linewidth]{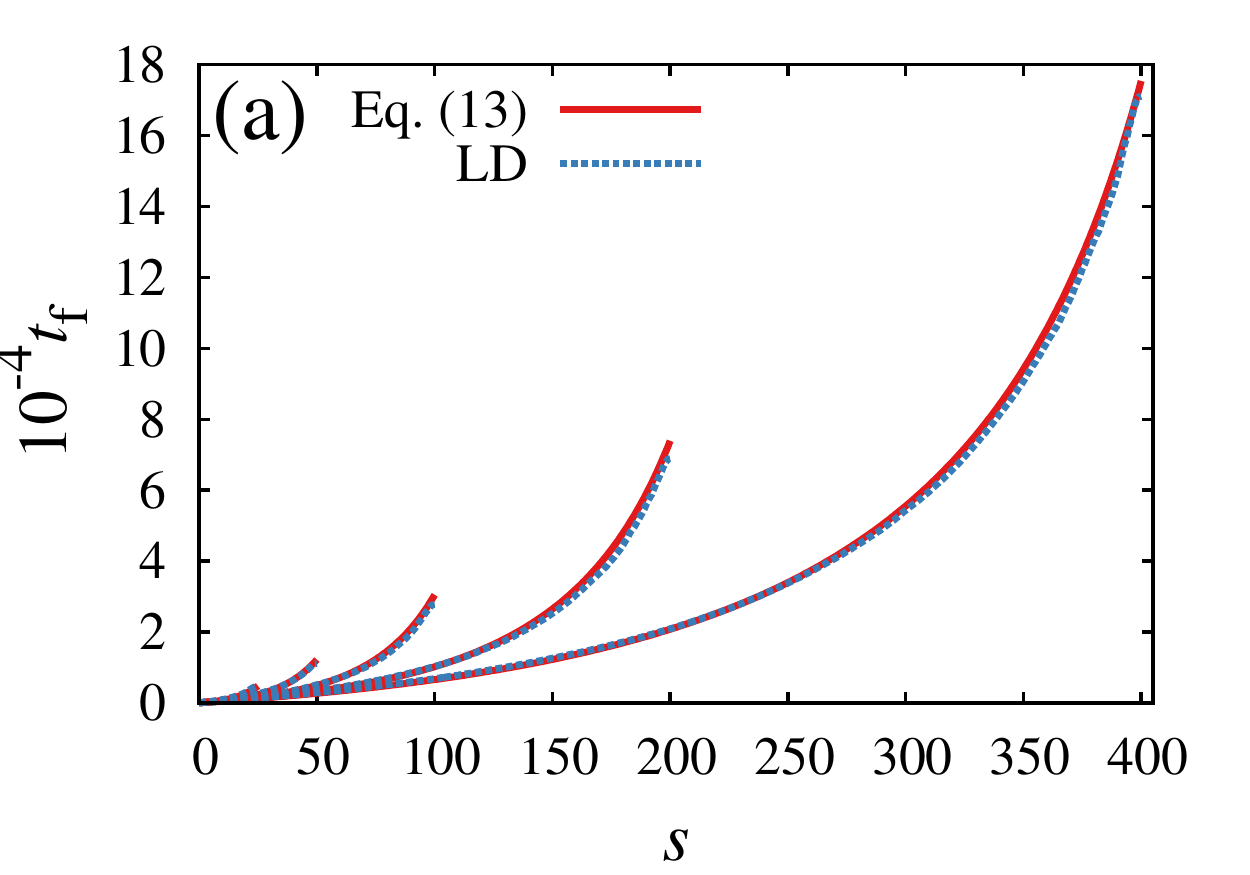}
\includegraphics[width=0.9\linewidth]{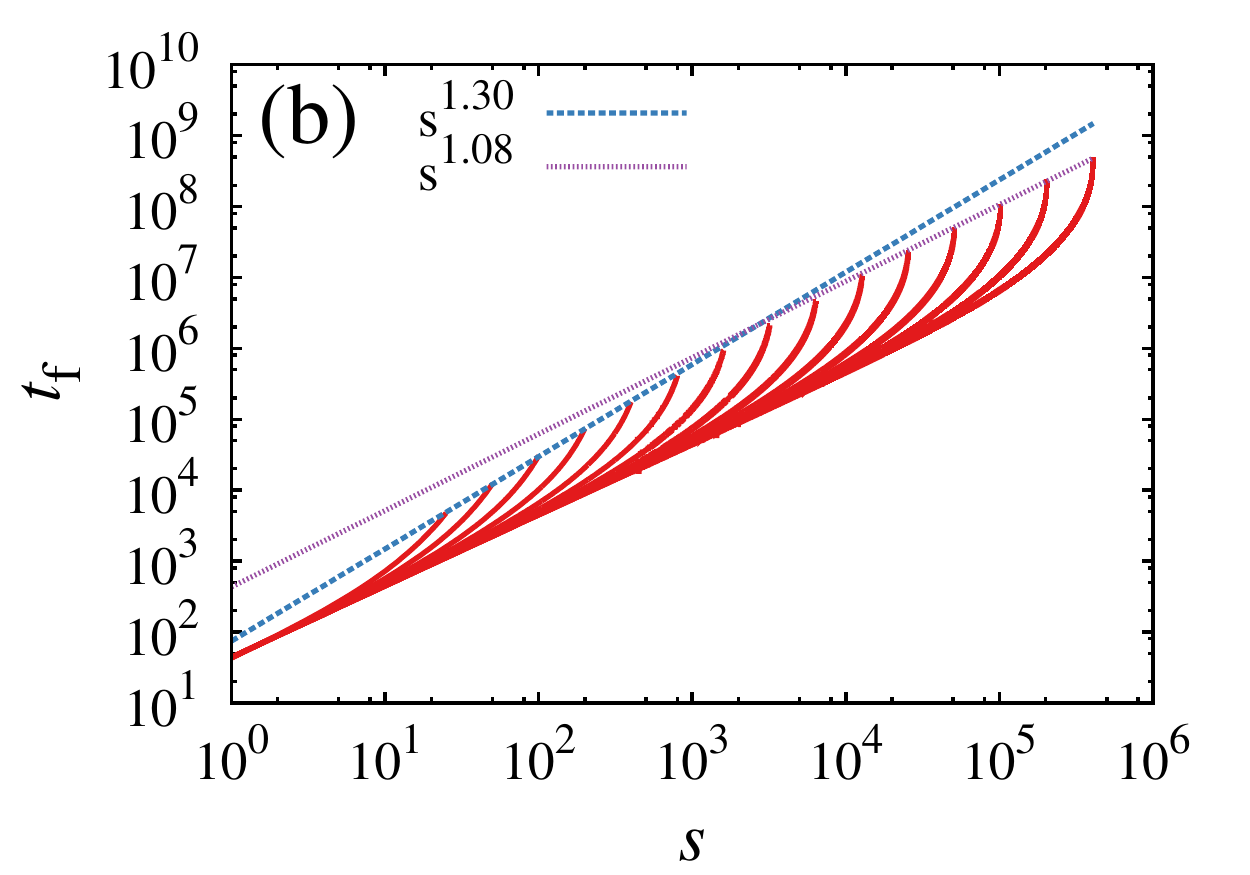}
\caption{(Color online) (a) First passage times numerically computed from Eq.~\eqref{equ:firstPassageTime} for $N_0 \leq 400$ (solid line) and the corresponding first passage times from LD simulations with $\rho_0=1.0$ (dotted line). (b) First passage times computed from Eq.~\eqref{equ:firstPassageTime} on a logarithmic scale for a large range of $N_0$. The endpoints seem to scale for small $N_0 \leq 400$ as $N_0^{1.3}$ while the apparent scaling tends towards linear with extremely high $N_0$. The dashed line shows a power law fit to $N_0 \leq 400$ while the dotted line is a power law fit to $N_0 > 1\times10^5$. We selected parameters as $\xi m = 254.5$ and $kT = 5.23$ as they were found to give a fairly good correspondence between the Langevin simulations and the model.}
\label{fig:fpt}
\end{figure}

In Ref.~\cite{piili_capsid} we derived an analytical estimate for the ejection time of the polymer under the assumption of a purely exponential pore force. However, the presented treatment is not accurate at the very final stage of the ejection if the force is assumed to decay to zero in the end as presented in Fig.~\ref{fig:rho}~(a). This is because the ejection stalls completely if the driving force vanishes. Thus, the ejection cannot complete without the help of diffusion. Also, as the ejection slows down considerably in the end, the final stage of ejection has a major effect on the total ejection time. We can estimate the first passage times $t_{\rm f}(s)$ based on the pore force solely by using the formula~\cite{gardiner_stochastic_methods}
\begin{align}\label{equ:firstPassageTime}
 t_f(s) = \frac{kT}{\xi m}\int_0^{s} \exp{\left( - \frac{U(y)}{kT} \right)} \int_y^{s} \exp{\left(\frac{U(z)}{kT}\right)} {\rm d}z\;{\rm d}y,
\end{align}
where
\begin{align}
U(s) = - \int_0^s f(s) {\rm d} s = \frac{C}{B}\left[N_0 \exp{\left(B \frac{N_0-s}{N_0}\right)} + Bs\right]
\end{align}
is the pore potential derived from the shifted exponential pore force $f(s)$ presented in Fig.~\ref{fig:rho}~(a). Here, we use $\rho$ instead of the effective density for simplicity. The integral is not analytically solvable, but by numerical computation we obtain a good correspondence with simulations in the regime $N_0\leq400$ as shown in Fig.~\ref{fig:fpt}~(a). The parameters $\xi m = 254.5$ and $k T = 5.23$ in Eq.~\eqref{equ:firstPassageTime} were chosen such that a good correspondence is obtained with all the presented curves, with the emphasis on the endpoints.

The parameter $\xi m$ describes the friction of both the solvent and of the pore while $k T$ describes the friction of the solvent multiplied by the temperature. Hence, we cannot expect to obtain a direct mapping of the parameters from the Langevin equation. The force parameter values $C = 0.411$ and $B = 2.70$ were obtained from a fit to the force measured for a polymer of length $N_0=200$ at the initial density $\rho_0=1.0$.

Figure~\ref{fig:fpt}~(b) shows the first passage time curves obtained from the numerical integration of Eq.~\eqref{equ:firstPassageTime} in the logarithmic scale. The endpoints of the curves reveal that the apparent scaling of $\tau$ vs. $N_0$ tends towards linear when $N_0$ grows extremely large. Therefore we can safely assume that the apparent scaling observed for $N_0 \leq 400$ indeed is a finite size effect even for a strictly vanishing pore force.

When numerically integrating Eq.~\eqref{equ:firstPassageTime} we observed that the obtained first passage times are extremely sensitive to the choice of the force. A force that decays to negative values for very small densities leads to exponentially growing ejection times. If the force remains above zero throughout the ejection, including the endpoint $s/N_0=1$, the formula leads to linear scaling already with very small $N_0$. If in the end of the ejection there is a finitely long regime in $s/N_0$ where the force is exactly zero, this diffusive region gives a scaling with the exponent 2. The presented model where force decays to zero exactly at the end of the ejection presents a special case where it is not evident beforehand what happens to the scaling. It also seems to best describe the simulation results received and explain why we obtain such a nice apparent scaling while everything seems to imply linear scaling for very long polymers.

\begin{figure}[b]
\includegraphics[width=\linewidth]{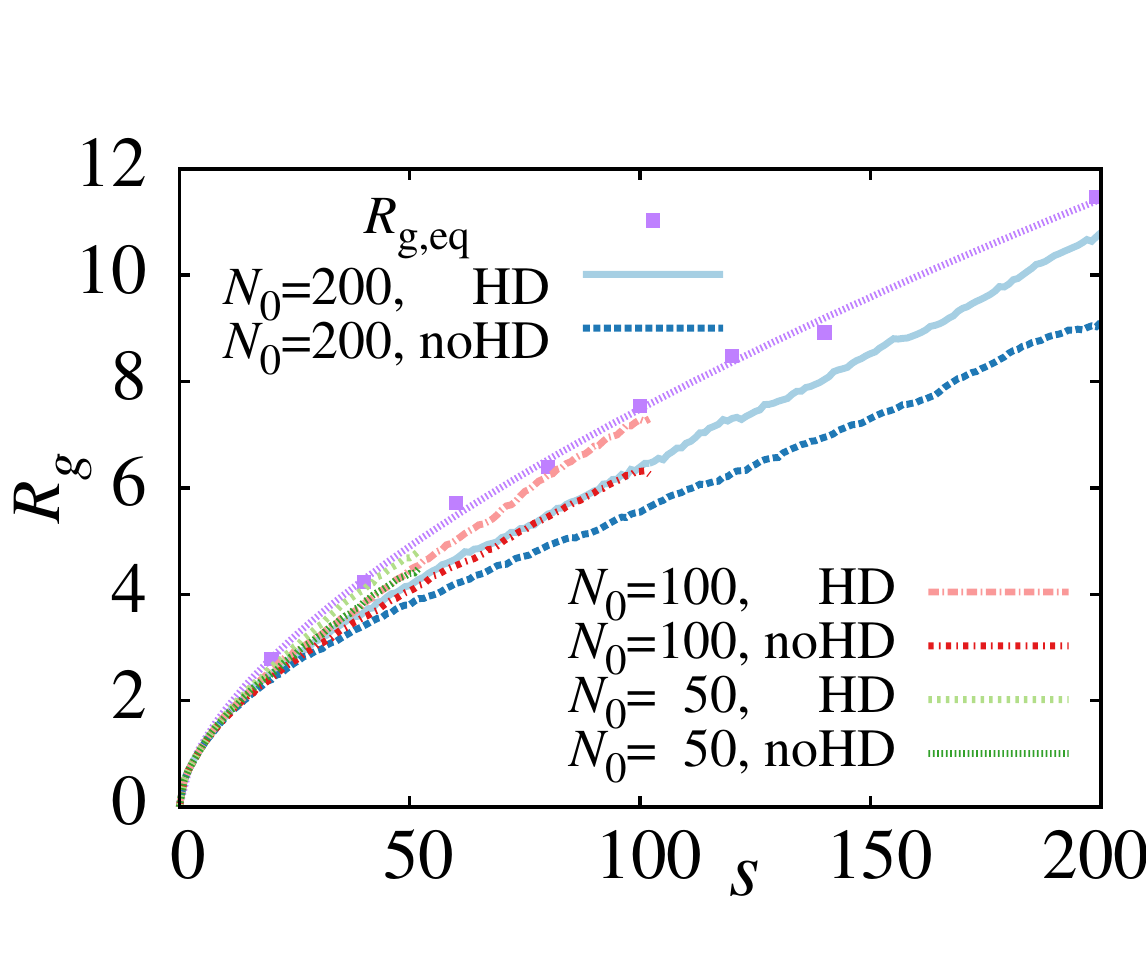}
\caption{(Color online) Radii of gyration $R_g$ outside the capsid during ejection as a function of the reaction coordinate $s$ with (HD) and without (noHD) hydrodynamics. Initial monomer density $\rho_0 = 1.0$ and polymer lengths $N_0 = 50,100, \; {\rm and } \; 200$ depicted. For reference, the corresponding radii of gyration in equilibrium $R_{g,\rm eq}$ are also depicted. They are measured from outside of a capsid of volume $V = 200/1.0$. With hydrodynamics included, the polymer $R_g$ is generally larger than without hydrodynamics, even though the ejection is faster with enabled hydrodynamics.}
\label{fig:rgOutDuringEjection}
\end{figure}

\subsection{The radius of gyration}\label{sec:rg_relaxation}

We have previously found that the radius of gyration of the polymer segment on the {\it trans} side grows roughly as $R_g \sim s^{0.6}$, which means that the ejecting polymer segment is not far from equilibrium~\cite{piili_capsid}, unlike in the case of driven translocation~\cite{lehtola_epl}. Here, using SRD we measure $R_g(s)$ for the ejected polymer segment with and without hydrodynamics and $R_{g,\rm eq}$ for polymers of different lengths $N = s$ at equilibrium, see Fig.~\ref{fig:rgOutDuringEjection}. It is seen that throughout the ejection the ejected polymer segment is slightly compressed compared to the equilibrium conformation, that is, $R_g < R_{g,\rm eq}$. Hence, the ejected polymer segment remains slightly out of equilibrium throughout the ejection. $R_g$ is larger and closer to $R_{g,\rm eq}$ when hydrodynamics is included. So, hydrodynamics speeds up the relaxation of a polymer even more than it speeds up the ejection. The same observation was made concerning driven polymer translocation~\cite{jaakko}.

The radius of gyration of the {\it trans} side polymer segment manifests clearly one characteristic of the polymer ejection that makes it impossible for the ejection waiting time to scale with the reaction coordinate $s$. $R_g$ is seen to be smaller for long than short polymers. In other words, long polymers are driven farther out of equilibrium than short polymers starting from the same initial monomer density. The monomer density $\rho$ inside the capsid decays faster with increasing $s$ for a short polymer than for a long polymer. For example, for a polymer of $N_0 = 50$ $\rho = 0$ at $s = 50$, but not so for a polymer of $N_0 = 100$. Instead, $\rho$ decays with $s/N_0$ identically for all $N_0$. So, since polymers of different lengths are driven by different force at a same $s$, there can be no universal scaling of the waiting time $t_w$ with $s$. For the same reason, the longer the polymer, the farther out of equilibrium it is for any given $s > 0$.

\label{fig:notrans_LD}
\subsection{Modified models}
\label{modmodels}
\begin{figure}[tb]
\includegraphics[width=.8\linewidth]{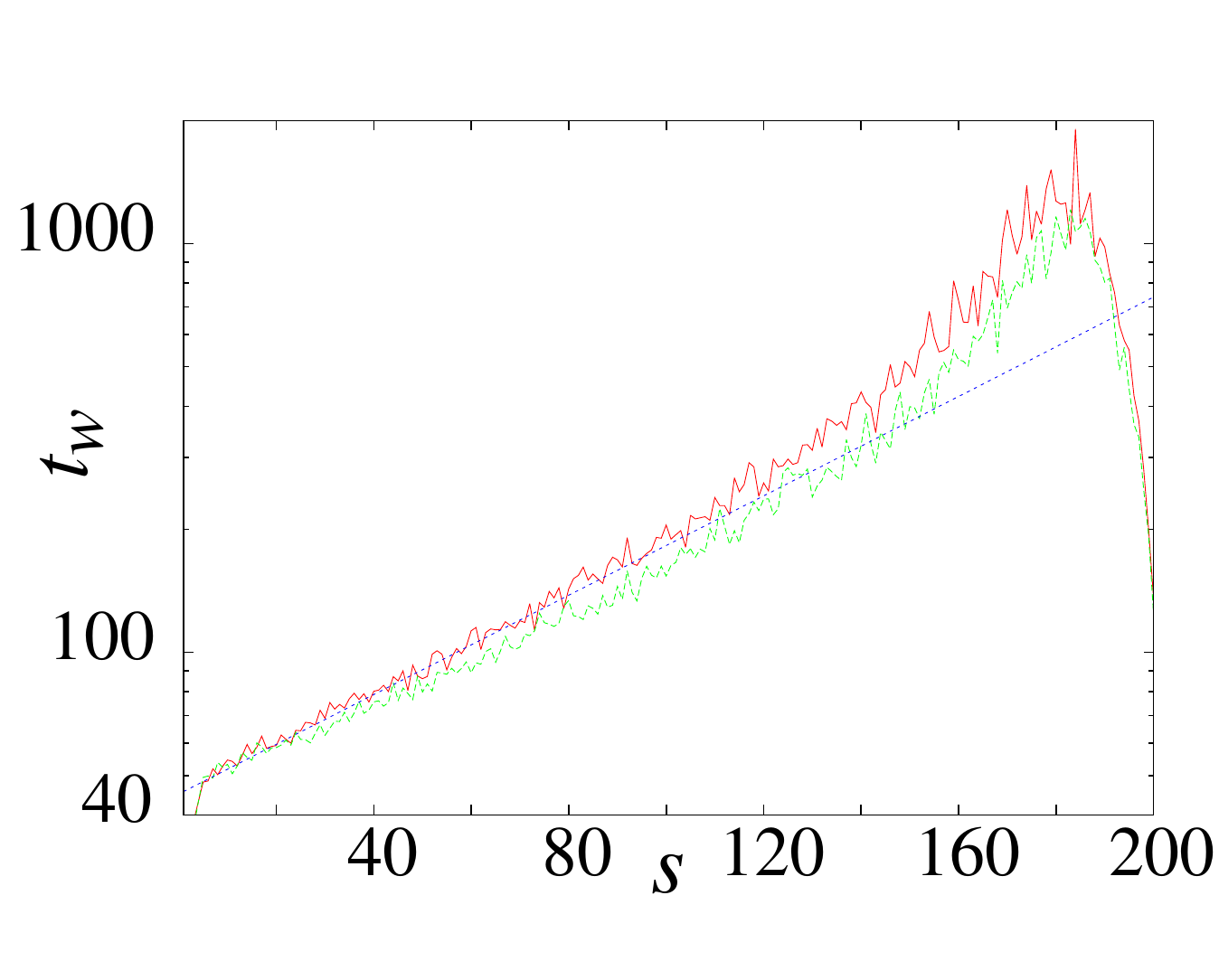}
\caption{(Color online) Waiting time $t_w(s)$ for the full polymer (solid line, above) and for a polymer where the beads on the \textit{trans} side are removed during ejection (dashed line, below) such that there are $N_{\textit{trans}}=5$ beads on the cis side. The straight line shows $t_w\sim\exp{\left((2.8/N_0) s\right)}$.}
\label{fig:notrans}
\end{figure}
From the above observations it is evident that the polymer ejection is inherently a non-equilibrium process whose non-equilibrium characteristics are more enhanced for long polymers. As stated in Section~\ref{sec:force}, the two non-equilibrium mechanisms potentially affecting the ejection dynamics are tension propagation on the {\it cis} side and crowding on the {\it trans} side. In the case of driven polymer translocation we have shown that crowding has no effect, but the translocation dynamics is in practice determined by the dynamics of tension propagation~\cite{pauli_translocation}. For a worm-like-chain in capsid ejection it has been shown that crowding slows down the ejection~\cite{lawati_capsid_tail}. This result, however, cannot be generalized to the flexible chain polymer model.

In order to determine the dominating non-equilibrium effect in the polymer ejection from a capsid we remove beads from the {\it trans} side during ejection. We simulate ejection of polymers of $N_0 = 200$ such that the maximum number of beads on the {\it trans} side at any moment is $N_{\rm trans} = 5$, $10$, or $160$. For $N_{\rm trans} = 160$ the waiting time profile $t_w(s)$ is identical to that of the full polymer ejection, where no beads are removed. $t_w(s)$ for $N_{\rm trans} = 5$ and $10$ deviate from $t_w(s)$ of full polymer, deviation being greater for $N_{\rm trans} = 5$. Fig.~\ref{fig:notrans} shows $t_w(s)$ for the full polymer and for $N_{\rm trans} = 5$. $t_w(s)$ for $N_{\rm trans} = 5$ is seen to be clearly smaller than $t_w(s)$ for the full polymer leading to the ejection time of 18\% smaller with $N_{\rm trans} = 5$ than with the full polymer. This can only be due to crowding not slowing down the ejection for $N_{\rm trans} = 5$. This is in contrast to driven translation where the effect of crowding is far less significant than the effect the tension propagation has on the dynamics. In translocation with $N_0=200$, the ejection time with $N_{\rm trans}=5$ is 7.9\% smaller with $f_d=2$ and 4.6\% smaller with $f_d=8$ than ejection times with full polymer. In the ejection process tension can significantly propagate only at the final stages of the process when the monomers are less densely packed. Also the force driving the polymer has decreased at this point, so tension propagation will be mild. Under these circumstances crowding, although being weaker than in the driven polymer translocation, shows more prominently in the resulting polymer ejection dynamics.

\section{Conclusion}\label{sec:con}
We have studied the ejection of a fully flexible polymer chain from a spherical capsid modeled using simplistic boundary conditions. Hydrodynamic interactions were simulated using stochastic rotation dynamics (SRD) coupled with molecular dynamics. We also used SRD with hydrodynamic interactions switched off to better pin down effects due to hydrodynamics. The less established SRD method was compared to the thoroughly understood Langevin dynamics (LD) to verify its suitability for simulations of this kind. Results obtained using SRD without hydrodynamics were found to be in good agreement with those obtained using the LD model whose friction was matched. Computational efficiency of LD allowed us to perform more precise force measurements than before.

From measured ejection times $\tau$ for different polymer lengths $N_0$ the apparent scaling $\tau \sim N_0^\beta$ was obtained for $N_0\leq400$. The differences in the fitted effective exponents $\beta$ obtained for different models make sense. Included hydrodynamic interactions not only reduce the ejection time but also reduce the $\beta$ exponent which is in line with results of forced translocation~\cite{jaakko, lehtola_epl}. This was addressed to hydrodynamic correlations reducing friction proportionately less in the pore region than outside it, which contributes towards linear scaling. We showed, however, that basing the analysis solely on ejection times, as in many previous studies, leads to an incorrect characterization of the process.

The waiting time $t_w(s)$ proved most valuable for understanding the polymer ejection dynamics. It describes how long it takes for the individual polymer bead $s$ to permanently exit the capsid after the final exit of the bead $s-1$. In the previous study we concluded that the waiting time $t_w(s)$ is of exponential form~\cite{piili_capsid}. The more precise measurements and analysis conducted here reveal that $t_w(s)$ is actually more accurately described by a sum of two exponentials. After about 63\% of the polymer has ejected, the ejection starts to slow down even more considerably and the second exponential starts to dominate the waiting time. When hydrodynamics is included $t_w(s)$ is of almost identical form as without hydrodynamics, only by a constant factor smaller in magnitude.

Remarkably, waiting times $t_w(s)$ for polymers of different lengths $N_0$ starting their ejection at the same initial monomer density $\rho_0$ collapse when plotted as a function of the normalized reaction coordinate $s/N_0$. This implies, by definition, that the ejection times should scale linearly with $N_0$. Only the final retraction of $t_w$ makes the scaling superlinear for polymers of moderate length. However, the effect of the final retraction becomes proportionately smaller for longer polymers, ultimately vanishing for very long polymers. Hence, linear scaling is approached for extensively long polymers and the cumulative waiting time is given by $t(s) = N_0 h(s/N_0)$, where $h(s/N_0)$ is a scaling function.

Our LD model allowed us to investigate the force exerted at the pore more carefully than in our previous study, due to its computational efficiency. We found that the position from where the force is measured has a large effect on the measurements. The force measured at the pore entrance is very accurately an exponential function of the monomer density inside the capsid. On the other hand, the force measured at the middle of the pore is exponential only for densities exceeding $\sim 0.4$. To capture the form of the force more precisely we need to reduce a constant factor from the exponential. This takes the force to zero for zero monomer density. This was observed to significantly affect the ejection times. We also found out that the force is a function of the effective density, where the effective radius of the inside of the capsid is $\sim 0.3$ larger than the real radius used in the simulations. This is due to the repulsive LJ potential overlapping with the capsid walls. This effect diminishes as the capsid volume increases.

By inserting the measured exponential form for the force, from which the offset term is reduced, into the first passage time formula for a random walk we are able to quite accurately reproduce the obtained first passage times from simulations. The numerical integration of the formula also reveals that this type of force leads to linear scaling for extremely large $N_0$.

The radius of gyration $R_g$ outside the capsid shows that the part of the polymer on the {\it trans} side is more compact than the corresponding polymer in equilibrium. When hydrodynamics is included the {\it trans} side is larger and therefore closer to equilibrium $R_g$ than without hydrodynamics. This occurs even though the faster ejection with hydrodynamics allows the polymer conformation less time to expand. In a modified model, where the number of monomers on the {\it trans} side are kept constant by continually removing them, a polymer ejects slightly faster than the corresponding full polymer. Hence, crowding has a small effect on ejection dynamics for the flexible polymer corroborating the findings from the measurements of $R_g$. Finally, hydrodynamics has a fairly weak effect on capsid ejection, including the crowding. Most importantly, it does not alter the universal form of the waiting time versus the number of ejected monomers. All evidence from our measurements goes to show that the apparent superlinear scaling of the  ejection time with the  polymer length $\tau \sim N_0^\beta$ tends to linear scaling for extremely long polymers.

\begin{acknowledgments}
The computational resources of CSC-IT Center for Science, Finland, and Aalto Science-IT project are acknowledged. The work of Joonas Piili was supported by The Emil Aaltonen Foundation and Finnish Foundation for Technology Promotion. The work of Pauli Suhonen is supported by The Emil Aaltonen Foundation.
\end{acknowledgments}

\bibliography{references.bib}

\begin{thebibliography}{35}%
\makeatletter
\providecommand \@ifxundefined [1]{%
 \@ifx{#1\undefined}
}%
\providecommand \@ifnum [1]{%
 \ifnum #1\expandafter \@firstoftwo
 \else \expandafter \@secondoftwo
 \fi
}%
\providecommand \@ifx [1]{%
 \ifx #1\expandafter \@firstoftwo
 \else \expandafter \@secondoftwo
 \fi
}%
\providecommand \natexlab [1]{#1}%
\providecommand \enquote  [1]{``#1''}%
\providecommand \bibnamefont  [1]{#1}%
\providecommand \bibfnamefont [1]{#1}%
\providecommand \citenamefont [1]{#1}%
\providecommand \href@noop [0]{\@secondoftwo}%
\providecommand \href [0]{\begingroup \@sanitize@url \@href}%
\providecommand \@href[1]{\@@startlink{#1}\@@href}%
\providecommand \@@href[1]{\endgroup#1\@@endlink}%
\providecommand \@sanitize@url [0]{\catcode `\\12\catcode `\$12\catcode
  `\&12\catcode `\#12\catcode `\^12\catcode `\_12\catcode `\%12\relax}%
\providecommand \@@startlink[1]{}%
\providecommand \@@endlink[0]{}%
\providecommand \url  [0]{\begingroup\@sanitize@url \@url }%
\providecommand \@url [1]{\endgroup\@href {#1}{\urlprefix }}%
\providecommand \urlprefix  [0]{URL }%
\providecommand \Eprint [0]{\href }%
\providecommand \doibase [0]{http://dx.doi.org/}%
\providecommand \selectlanguage [0]{\@gobble}%
\providecommand \bibinfo  [0]{\@secondoftwo}%
\providecommand \bibfield  [0]{\@secondoftwo}%
\providecommand \translation [1]{[#1]}%
\providecommand \BibitemOpen [0]{}%
\providecommand \bibitemStop [0]{}%
\providecommand \bibitemNoStop [0]{.\EOS\space}%
\providecommand \EOS [0]{\spacefactor3000\relax}%
\providecommand \BibitemShut  [1]{\csname bibitem#1\endcsname}%
\let\auto@bib@innerbib\@empty
\bibitem [{\citenamefont {Glasgow}\ and\ \citenamefont
  {Tullman-Ercek}(2014)}]{glasgow}%
  \BibitemOpen
  \bibfield  {author} {\bibinfo {author} {\bibfnamefont {J.}~\bibnamefont
  {Glasgow}}\ and\ \bibinfo {author} {\bibfnamefont {D.}~\bibnamefont
  {Tullman-Ercek}},\ }\bibfield  {title} {\enquote {\bibinfo {title}
  {Production and applications of engineered viral capsids},}\ }\href@noop {}
  {\bibfield  {journal} {\bibinfo  {journal} {Appl. Microbiol. Biotechnol.}\
  }\textbf {\bibinfo {volume} {98}},\ \bibinfo {pages} {5847} (\bibinfo {year}
  {2014})}\BibitemShut {NoStop}%
\bibitem [{\citenamefont {Muthukumar}(2001)}]{muthukumar1}%
  \BibitemOpen
  \bibfield  {author} {\bibinfo {author} {\bibfnamefont {M.}~\bibnamefont
  {Muthukumar}},\ }\bibfield  {title} {\enquote {\bibinfo {title}
  {Translocation of a confined polymer through a hole},}\ }\href {\doibase
  10.1103/PhysRevLett.86.3188} {\bibfield  {journal} {\bibinfo  {journal}
  {Phys. Rev. Lett.}\ }\textbf {\bibinfo {volume} {86}},\ \bibinfo {pages}
  {3188} (\bibinfo {year} {2001})}\BibitemShut {NoStop}%
\bibitem [{\citenamefont {Forrey}\ and\ \citenamefont
  {Muthukumar}(2006)}]{muthukumar2}%
  \BibitemOpen
  \bibfield  {author} {\bibinfo {author} {\bibfnamefont {Christopher}\
  \bibnamefont {Forrey}}\ and\ \bibinfo {author} {\bibfnamefont
  {M.}~\bibnamefont {Muthukumar}},\ }\bibfield  {title} {\enquote {\bibinfo
  {title} {Langevin dynamics simulations of genome packing in bacteriophage},}\
  }\href {\doibase http://dx.doi.org/10.1529/biophysj.105.073429} {\bibfield
  {journal} {\bibinfo  {journal} {Biophys. J.}\ }\textbf {\bibinfo {volume}
  {91}},\ \bibinfo {pages} {25} (\bibinfo {year} {2006})}\BibitemShut {NoStop}%
\bibitem [{\citenamefont {Smith}\ \emph {et~al.}(2001)\citenamefont {Smith},
  \citenamefont {Tans}, \citenamefont {Smith}, \citenamefont {Andersen},\ and\
  \citenamefont {Bustamante}}]{smith}%
  \BibitemOpen
  \bibfield  {author} {\bibinfo {author} {\bibfnamefont {D.~E.}\ \bibnamefont
  {Smith}}, \bibinfo {author} {\bibfnamefont {S.~B.}\ \bibnamefont {Tans}},
  \bibinfo {author} {\bibfnamefont {S.}~\bibnamefont {Smith}, \bibfnamefont
  {S.~B.~Grimes}}, \bibinfo {author} {\bibfnamefont {D.~L.}\ \bibnamefont
  {Andersen}}, \ and\ \bibinfo {author} {\bibfnamefont {C.}~\bibnamefont
  {Bustamante}},\ }\bibfield  {title} {\enquote {\bibinfo {title} {The
  bacteriophage $\phi$29 portal motor can package {DNA} against a large
  internal force},}\ }\href@noop {} {\bibfield  {journal} {\bibinfo  {journal}
  {Nature}\ }\textbf {\bibinfo {volume} {413}},\ \bibinfo {pages} {748}
  (\bibinfo {year} {2001})}\BibitemShut {NoStop}%
\bibitem [{\citenamefont {Grayson}\ \emph {et~al.}(2007)\citenamefont
  {Grayson}, \citenamefont {Han}, \citenamefont {Winther},\ and\ \citenamefont
  {Phillips}}]{grayson}%
  \BibitemOpen
  \bibfield  {author} {\bibinfo {author} {\bibfnamefont {Paul}\ \bibnamefont
  {Grayson}}, \bibinfo {author} {\bibfnamefont {Lin}\ \bibnamefont {Han}},
  \bibinfo {author} {\bibfnamefont {Tabita}\ \bibnamefont {Winther}}, \ and\
  \bibinfo {author} {\bibfnamefont {Rob}\ \bibnamefont {Phillips}},\ }\bibfield
   {title} {\enquote {\bibinfo {title} {Real-time observations of single
  bacteriophage $\lambda$ {DNA} ejections in vitro},}\ }\href@noop {}
  {\bibfield  {journal} {\bibinfo  {journal} {Proc. Natl. Acad. Sci}\ }\textbf
  {\bibinfo {volume} {104}},\ \bibinfo {pages} {14652} (\bibinfo {year}
  {2007})}\BibitemShut {NoStop}%
\bibitem [{\citenamefont {Ali}\ \emph {et~al.}(2006)\citenamefont {Ali},
  \citenamefont {Marenduzzo},\ and\ \citenamefont {Yeomans}}]{ali}%
  \BibitemOpen
  \bibfield  {author} {\bibinfo {author} {\bibfnamefont {I.}~\bibnamefont
  {Ali}}, \bibinfo {author} {\bibfnamefont {D.}~\bibnamefont {Marenduzzo}}, \
  and\ \bibinfo {author} {\bibfnamefont {J.~M.}\ \bibnamefont {Yeomans}},\
  }\bibfield  {title} {\enquote {\bibinfo {title} {Polymer packaging and
  ejection in viral capsids: Shape matters},}\ }\href@noop {} {\bibfield
  {journal} {\bibinfo  {journal} {Phys. Rev. Lett.}\ }\textbf {\bibinfo
  {volume} {96}},\ \bibinfo {pages} {208102} (\bibinfo {year}
  {2006})}\BibitemShut {NoStop}%
\bibitem [{\citenamefont {Ghosal}(2012)}]{ghosal}%
  \BibitemOpen
  \bibfield  {author} {\bibinfo {author} {\bibfnamefont {Sandip}\ \bibnamefont
  {Ghosal}},\ }\bibfield  {title} {\enquote {\bibinfo {title} {Capstan friction
  model for {DNA} ejection from bacteriophages},}\ }\href {\doibase
  10.1103/PhysRevLett.109.248105} {\bibfield  {journal} {\bibinfo  {journal}
  {Phys. Rev. Lett.}\ }\textbf {\bibinfo {volume} {109}},\ \bibinfo {pages}
  {248105} (\bibinfo {year} {2012})}\BibitemShut {NoStop}%
\bibitem [{\citenamefont {Cacciuto}\ and\ \citenamefont
  {Luijten}(2006{\natexlab{a}})}]{cacciuto2}%
  \BibitemOpen
  \bibfield  {author} {\bibinfo {author} {\bibfnamefont {A.}~\bibnamefont
  {Cacciuto}}\ and\ \bibinfo {author} {\bibfnamefont {E.}~\bibnamefont
  {Luijten}},\ }\bibfield  {title} {\enquote {\bibinfo {title}
  {Confinement-driven translocation of a flexible polymer},}\ }\href@noop {}
  {\bibfield  {journal} {\bibinfo  {journal} {Phys. Rev. Lett.}\ }\textbf
  {\bibinfo {volume} {96}},\ \bibinfo {pages} {238104} (\bibinfo {year}
  {2006}{\natexlab{a}})}\BibitemShut {NoStop}%
\bibitem [{\citenamefont {Sakaue}\ and\ \citenamefont
  {Yoshinaga}(2009)}]{sakaue_polymer_decompression}%
  \BibitemOpen
  \bibfield  {author} {\bibinfo {author} {\bibfnamefont {Takahiro}\
  \bibnamefont {Sakaue}}\ and\ \bibinfo {author} {\bibfnamefont {Natsuhiko}\
  \bibnamefont {Yoshinaga}},\ }\bibfield  {title} {\enquote {\bibinfo {title}
  {Dynamics of polymer decompression: Expansion, unfolding, and ejection},}\
  }\href {\doibase 10.1103/PhysRevLett.102.148302} {\bibfield  {journal}
  {\bibinfo  {journal} {Phys. Rev. Lett.}\ }\textbf {\bibinfo {volume} {102}},\
  \bibinfo {pages} {148302} (\bibinfo {year} {2009})}\BibitemShut {NoStop}%
\bibitem [{\citenamefont {Linna}\ \emph {et~al.}(2014)\citenamefont {Linna},
  \citenamefont {Moisio}, \citenamefont {Suhonen},\ and\ \citenamefont
  {Kaski}}]{riku_dynamics_of_ejection}%
  \BibitemOpen
  \bibfield  {author} {\bibinfo {author} {\bibfnamefont {R.~P.}\ \bibnamefont
  {Linna}}, \bibinfo {author} {\bibfnamefont {J.~E.}\ \bibnamefont {Moisio}},
  \bibinfo {author} {\bibfnamefont {P.~M.}\ \bibnamefont {Suhonen}}, \ and\
  \bibinfo {author} {\bibfnamefont {K.}~\bibnamefont {Kaski}},\ }\bibfield
  {title} {\enquote {\bibinfo {title} {Dynamics of polymer ejection from
  capsid},}\ }\href {\doibase 10.1103/PhysRevE.89.052702} {\bibfield  {journal}
  {\bibinfo  {journal} {Phys. Rev. E}\ }\textbf {\bibinfo {volume} {89}},\
  \bibinfo {pages} {052702} (\bibinfo {year} {2014})}\BibitemShut {NoStop}%
\bibitem [{\citenamefont {Malevanets}\ and\ \citenamefont
  {Kapral}(1999)}]{malevanets_orig}%
  \BibitemOpen
  \bibfield  {author} {\bibinfo {author} {\bibfnamefont {Anatoly}\ \bibnamefont
  {Malevanets}}\ and\ \bibinfo {author} {\bibfnamefont {Raymond}\ \bibnamefont
  {Kapral}},\ }\bibfield  {title} {\enquote {\bibinfo {title} {Mesoscopic model
  for solvent dynamics},}\ }\href@noop {} {\bibfield  {journal} {\bibinfo
  {journal} {J. Chem. Phys.}\ }\textbf {\bibinfo {volume} {110}},\ \bibinfo
  {pages} {8605} (\bibinfo {year} {1999})}\BibitemShut {NoStop}%
\bibitem [{\citenamefont {Malevanets}\ and\ \citenamefont
  {Kapral}(2004)}]{malevanets_mesoscopic}%
  \BibitemOpen
  \bibfield  {author} {\bibinfo {author} {\bibfnamefont {Anatoly}\ \bibnamefont
  {Malevanets}}\ and\ \bibinfo {author} {\bibfnamefont {Raymond}\ \bibnamefont
  {Kapral}},\ }\bibfield  {title} {\enquote {\bibinfo {title} {Mesoscopic
  multi-particle collision model for fluid flow and molecular dynamics},}\
  }\href {http://www.springerlink.com/index/RWVHW98J9QE85FRU.pdf} {\bibfield
  {journal} {\bibinfo  {journal} {Novel Methods in Soft Matter Simulations}\
  }\textbf {\bibinfo {volume} {149}},\ \bibinfo {pages} {2258} (\bibinfo {year}
  {2004})}\BibitemShut {NoStop}%
\bibitem [{\citenamefont {Piili}\ and\ \citenamefont
  {Linna}(2015)}]{piili_capsid}%
  \BibitemOpen
  \bibfield  {author} {\bibinfo {author} {\bibfnamefont {J.}~\bibnamefont
  {Piili}}\ and\ \bibinfo {author} {\bibfnamefont {R.~P.}\ \bibnamefont
  {Linna}},\ }\bibfield  {title} {\enquote {\bibinfo {title} {Polymer ejection
  from strong spherical confinement},}\ }\href {\doibase
  10.1103/PhysRevE.92.062715} {\bibfield  {journal} {\bibinfo  {journal} {Phys.
  Rev. E}\ }\textbf {\bibinfo {volume} {92}},\ \bibinfo {pages} {062715}
  (\bibinfo {year} {2015})}\BibitemShut {NoStop}%
\bibitem [{\citenamefont {Allen}\ and\ \citenamefont
  {Tildesley}(2006)}]{allen}%
  \BibitemOpen
  \bibfield  {author} {\bibinfo {author} {\bibfnamefont {M.~P.}\ \bibnamefont
  {Allen}}\ and\ \bibinfo {author} {\bibfnamefont {D.~J.}\ \bibnamefont
  {Tildesley}},\ }\href@noop {} {\emph {\bibinfo {title} {Computer Simulation
  of Liquids}}}\ (\bibinfo  {publisher} {Clarendon Press},\ \bibinfo {address}
  {Oxford},\ \bibinfo {year} {2006})\BibitemShut {NoStop}%
\bibitem [{\citenamefont {Humphrey}\ \emph {et~al.}(1996)\citenamefont
  {Humphrey}, \citenamefont {Dalke},\ and\ \citenamefont {Schulten}}]{vmd}%
  \BibitemOpen
  \bibfield  {author} {\bibinfo {author} {\bibfnamefont {William}\ \bibnamefont
  {Humphrey}}, \bibinfo {author} {\bibfnamefont {Andrew}\ \bibnamefont
  {Dalke}}, \ and\ \bibinfo {author} {\bibfnamefont {Klaus}\ \bibnamefont
  {Schulten}},\ }\bibfield  {title} {\enquote {\bibinfo {title} {{VMD} --
  {V}isual {M}olecular {D}ynamics},}\ }\href@noop {} {\bibfield  {journal}
  {\bibinfo  {journal} {Journal of Molecular Graphics}\ }\textbf {\bibinfo
  {volume} {14}},\ \bibinfo {pages} {33} (\bibinfo {year} {1996})}\BibitemShut
  {NoStop}%
\bibitem [{\citenamefont {{Persistence of Vision Pty. Ltd.}}(2004)}]{povray}%
  \BibitemOpen
  \bibfield  {author} {\bibinfo {author} {\bibnamefont {{Persistence of Vision
  Pty. Ltd.}}},\ }\href@noop {} {} (\bibinfo {year} {2004}),\ \bibinfo {note}
  {{Persistence of Vision Raytracer Software, www.povray.org}}\BibitemShut
  {NoStop}%
\bibitem [{\citenamefont {Lamura}\ \emph {et~al.}(2001)\citenamefont {Lamura},
  \citenamefont {Gompper}, \citenamefont {Ihle},\ and\ \citenamefont
  {Kroll}}]{lamura_srd_poseuille}%
  \BibitemOpen
  \bibfield  {author} {\bibinfo {author} {\bibfnamefont {A.}~\bibnamefont
  {Lamura}}, \bibinfo {author} {\bibfnamefont {G.}~\bibnamefont {Gompper}},
  \bibinfo {author} {\bibfnamefont {T.}~\bibnamefont {Ihle}}, \ and\ \bibinfo
  {author} {\bibfnamefont {D.~M.}\ \bibnamefont {Kroll}},\ }\bibfield  {title}
  {\enquote {\bibinfo {title} {Multi-particle collision dynamics: Flow around a
  circular and a square cylinder},}\ }\href
  {http://stacks.iop.org/0295-5075/56/i=3/a=319} {\bibfield  {journal}
  {\bibinfo  {journal} {EPL (Europhysics Letters)}\ }\textbf {\bibinfo {volume}
  {56}},\ \bibinfo {pages} {319} (\bibinfo {year} {2001})}\BibitemShut
  {NoStop}%
\bibitem [{\citenamefont {Frenkel}\ and\ \citenamefont
  {Smit}(2001)}]{frenkel_moldy}%
  \BibitemOpen
  \bibfield  {author} {\bibinfo {author} {\bibfnamefont {D.}~\bibnamefont
  {Frenkel}}\ and\ \bibinfo {author} {\bibfnamefont {B.}~\bibnamefont {Smit}},\
  }\href@noop {} {\emph {\bibinfo {title} {Understanding molecular simulation:
  from algorithms to applications}}}\ (\bibinfo  {publisher} {Academic Press},\
  \bibinfo {year} {2001})\BibitemShut {NoStop}%
\bibitem [{\citenamefont {Kikuchi}\ \emph {et~al.}(2003)\citenamefont
  {Kikuchi}, \citenamefont {Pooley}, \citenamefont {Ryder},\ and\ \citenamefont
  {Yeomans}}]{kikuchi_transport}%
  \BibitemOpen
  \bibfield  {author} {\bibinfo {author} {\bibfnamefont {N.}~\bibnamefont
  {Kikuchi}}, \bibinfo {author} {\bibfnamefont {C.~M.}\ \bibnamefont {Pooley}},
  \bibinfo {author} {\bibfnamefont {J.~F.}\ \bibnamefont {Ryder}}, \ and\
  \bibinfo {author} {\bibfnamefont {J.~M.}\ \bibnamefont {Yeomans}},\
  }\bibfield  {title} {\enquote {\bibinfo {title} {Transport coefficients of a
  mesoscopic fluid dynamics model},}\ }\href {\doibase
  http://dx.doi.org/10.1063/1.1603721} {\bibfield  {journal} {\bibinfo
  {journal} {The Journal of Chemical Physics}\ }\textbf {\bibinfo {volume}
  {119}},\ \bibinfo {pages} {6388} (\bibinfo {year} {2003})}\BibitemShut
  {NoStop}%
\bibitem [{\citenamefont {Ihle}\ and\ \citenamefont
  {Kroll}(2001)}]{ihle_galilean}%
  \BibitemOpen
  \bibfield  {author} {\bibinfo {author} {\bibfnamefont {T.}~\bibnamefont
  {Ihle}}\ and\ \bibinfo {author} {\bibfnamefont {D.~M.}\ \bibnamefont
  {Kroll}},\ }\bibfield  {title} {\enquote {\bibinfo {title} {Stochastic
  rotation dynamics: A galilean-invariant mesoscopic model for fluid flow},}\
  }\href {\doibase 10.1103/PhysRevE.63.020201} {\bibfield  {journal} {\bibinfo
  {journal} {Phys. Rev. E}\ }\textbf {\bibinfo {volume} {63}},\ \bibinfo
  {pages} {020201} (\bibinfo {year} {2001})}\BibitemShut {NoStop}%
\bibitem [{\citenamefont {Wyvill}\ and\ \citenamefont
  {Kunii}(1985)}]{wyvill_csg}%
  \BibitemOpen
  \bibfield  {author} {\bibinfo {author} {\bibfnamefont {G.}~\bibnamefont
  {Wyvill}}\ and\ \bibinfo {author} {\bibfnamefont {L.}~\bibnamefont {Kunii},
  \bibfnamefont {T.}},\ }\bibfield  {title} {\enquote {\bibinfo {title} {{A
  functional model for constructive solid geometry}},}\ }\href@noop {}
  {\bibfield  {journal} {\bibinfo  {journal} {The Visual Computer}\ }\textbf
  {\bibinfo {volume} {1}},\ \bibinfo {pages} {3} (\bibinfo {year}
  {1985})}\BibitemShut {NoStop}%
\bibitem [{\citenamefont {Matthews}\ \emph {et~al.}(2009)\citenamefont
  {Matthews}, \citenamefont {Louis},\ and\ \citenamefont
  {Yeomans}}]{matthews_knot_ejection}%
  \BibitemOpen
  \bibfield  {author} {\bibinfo {author} {\bibfnamefont {Richard}\ \bibnamefont
  {Matthews}}, \bibinfo {author} {\bibfnamefont {A.~A.}\ \bibnamefont {Louis}},
  \ and\ \bibinfo {author} {\bibfnamefont {J.~M.}\ \bibnamefont {Yeomans}},\
  }\bibfield  {title} {\enquote {\bibinfo {title} {Knot-controlled ejection of
  a polymer from a virus capsid},}\ }\href {\doibase
  10.1103/PhysRevLett.102.088101} {\bibfield  {journal} {\bibinfo  {journal}
  {Phys. Rev. Lett.}\ }\textbf {\bibinfo {volume} {102}},\ \bibinfo {pages}
  {088101} (\bibinfo {year} {2009})}\BibitemShut {NoStop}%
\bibitem [{\citenamefont {Marenduzzo}\ \emph {et~al.}(2013)\citenamefont
  {Marenduzzo}, \citenamefont {Micheletti}, \citenamefont {Orlandini},\ and\
  \citenamefont {Sumners}}]{marenduzzo_topological_dna_ejection}%
  \BibitemOpen
  \bibfield  {author} {\bibinfo {author} {\bibfnamefont {Davide}\ \bibnamefont
  {Marenduzzo}}, \bibinfo {author} {\bibfnamefont {Cristian}\ \bibnamefont
  {Micheletti}}, \bibinfo {author} {\bibfnamefont {Enzo}\ \bibnamefont
  {Orlandini}}, \ and\ \bibinfo {author} {\bibfnamefont {De~Witt}\ \bibnamefont
  {Sumners}},\ }\bibfield  {title} {\enquote {\bibinfo {title} {Topological
  friction strongly affects viral {DNA} ejection},}\ }\href {\doibase
  10.1073/pnas.1306601110} {\bibfield  {journal} {\bibinfo  {journal}
  {Proceedings of the National Academy of Sciences}\ }\textbf {\bibinfo
  {volume} {110}},\ \bibinfo {pages} {20081--20086} (\bibinfo {year}
  {2013})}\BibitemShut {NoStop}%
\bibitem [{ben()}]{bending_potential}%
  \BibitemOpen
  \href@noop {} {}\bibinfo {note} {The form of bending potential used during
  packing to test the effect of initial conformation to ejection times is of
  the form\\$U_{\rm bend} = -\kappa \sum_i \left( \mathbf{r}_{i+1} -
  \mathbf{r}_{i} \right)\cdot\left( \mathbf{r}_{i} - \mathbf{r}_{i-1} \right)$,
  where $\kappa=20$}\BibitemShut {NoStop}%
\bibitem [{\citenamefont {Ermak}\ and\ \citenamefont {Buckholz}(1980)}]{ermak}%
  \BibitemOpen
  \bibfield  {author} {\bibinfo {author} {\bibfnamefont {Donald~L.}\
  \bibnamefont {Ermak}}\ and\ \bibinfo {author} {\bibfnamefont {Helen}\
  \bibnamefont {Buckholz}},\ }\bibfield  {title} {\enquote {\bibinfo {title}
  {Numerical integration of the langevin equation: Monte carlo simulation},}\
  }\href {\doibase http://dx.doi.org/10.1016/0021-9991(80)90084-4} {\bibfield
  {journal} {\bibinfo  {journal} {Journal of Computational Physics}\ }\textbf
  {\bibinfo {volume} {35}},\ \bibinfo {pages} {169} (\bibinfo {year}
  {1980})}\BibitemShut {NoStop}%
\bibitem [{\citenamefont {Doi}(1996)}]{doi_introduction_to_polymer_physics}%
  \BibitemOpen
  \bibfield  {author} {\bibinfo {author} {\bibfnamefont {Masao}\ \bibnamefont
  {Doi}},\ }\href@noop {} {\emph {\bibinfo {title} {Introduction to polymer
  physics}}}\ (\bibinfo  {publisher} {Oxford university press},\ \bibinfo
  {year} {1996})\BibitemShut {NoStop}%
\bibitem [{\citenamefont {Tinland}\ \emph {et~al.}(1997)\citenamefont
  {Tinland}, \citenamefont {Pluen}, \citenamefont {Sturm},\ and\ \citenamefont
  {Weill}}]{tinland_persistence_length}%
  \BibitemOpen
  \bibfield  {author} {\bibinfo {author} {\bibfnamefont {Bernard}\ \bibnamefont
  {Tinland}}, \bibinfo {author} {\bibfnamefont {Alain}\ \bibnamefont {Pluen}},
  \bibinfo {author} {\bibfnamefont {Jean}\ \bibnamefont {Sturm}}, \ and\
  \bibinfo {author} {\bibfnamefont {Gilbert}\ \bibnamefont {Weill}},\
  }\bibfield  {title} {\enquote {\bibinfo {title} {Persistence length of
  single-stranded {DNA}},}\ }\href {\doibase 10.1021/ma970381+} {\bibfield
  {journal} {\bibinfo  {journal} {Macromolecules}\ }\textbf {\bibinfo {volume}
  {30}},\ \bibinfo {pages} {5763} (\bibinfo {year} {1997})}\BibitemShut
  {NoStop}%
\bibitem [{\citenamefont {Manning}(2006)}]{manning2006persistence}%
  \BibitemOpen
  \bibfield  {author} {\bibinfo {author} {\bibfnamefont {Gerald~S}\
  \bibnamefont {Manning}},\ }\bibfield  {title} {\enquote {\bibinfo {title}
  {The persistence length of {DNA} is reached from the persistence length of
  its null isomer through an internal electrostatic stretching force},}\
  }\href@noop {} {\bibfield  {journal} {\bibinfo  {journal} {Biophysical
  journal}\ }\textbf {\bibinfo {volume} {91}},\ \bibinfo {pages} {3607}
  (\bibinfo {year} {2006})}\BibitemShut {NoStop}%
\bibitem [{\citenamefont {Rechendorff}\ \emph {et~al.}(2009)\citenamefont
  {Rechendorff}, \citenamefont {Witz}, \citenamefont {Adamcik},\ and\
  \citenamefont {Dietler}}]{rechendorff_length_per_bead}%
  \BibitemOpen
  \bibfield  {author} {\bibinfo {author} {\bibfnamefont {Kristian}\
  \bibnamefont {Rechendorff}}, \bibinfo {author} {\bibfnamefont {Guillaume}\
  \bibnamefont {Witz}}, \bibinfo {author} {\bibfnamefont {Jozef}\ \bibnamefont
  {Adamcik}}, \ and\ \bibinfo {author} {\bibfnamefont {Giovanni}\ \bibnamefont
  {Dietler}},\ }\bibfield  {title} {\enquote {\bibinfo {title} {Persistence
  length and scaling properties of single-stranded {DNA} adsorbed on modified
  graphite},}\ }\href@noop {} {\bibfield  {journal} {\bibinfo  {journal} {The
  Journal of chemical physics}\ }\textbf {\bibinfo {volume} {131}},\ \bibinfo
  {pages} {095103} (\bibinfo {year} {2009})}\BibitemShut {NoStop}%
\bibitem [{\citenamefont {Lehtola}\ \emph {et~al.}(2009)\citenamefont
  {Lehtola}, \citenamefont {Linna},\ and\ \citenamefont {Kaski}}]{lehtola_epl}%
  \BibitemOpen
  \bibfield  {author} {\bibinfo {author} {\bibfnamefont {V.~V.}\ \bibnamefont
  {Lehtola}}, \bibinfo {author} {\bibfnamefont {R.~P.}\ \bibnamefont {Linna}},
  \ and\ \bibinfo {author} {\bibfnamefont {K.}~\bibnamefont {Kaski}},\
  }\bibfield  {title} {\enquote {\bibinfo {title} {Dynamics of forced
  biopolymer translocation},}\ }\href
  {http://stacks.iop.org/0295-5075/85/i=5/a=58006} {\bibfield  {journal}
  {\bibinfo  {journal} {EPL (Europhysics Letters)}\ }\textbf {\bibinfo {volume}
  {85}},\ \bibinfo {pages} {58006} (\bibinfo {year} {2009})}\BibitemShut
  {NoStop}%
\bibitem [{\citenamefont {Moisio}\ \emph {et~al.}(2016)\citenamefont {Moisio},
  \citenamefont {Piili},\ and\ \citenamefont {Linna}}]{jaakko}%
  \BibitemOpen
  \bibfield  {author} {\bibinfo {author} {\bibfnamefont {J.~E.}\ \bibnamefont
  {Moisio}}, \bibinfo {author} {\bibfnamefont {J.}~\bibnamefont {Piili}}, \
  and\ \bibinfo {author} {\bibfnamefont {R.~P.}\ \bibnamefont {Linna}},\
  }\bibfield  {title} {\enquote {\bibinfo {title} {Driven polymer translocation
  in good and bad solvent: Effects of hydrodynamics and tension propagation},}\
  }\href {\doibase 10.1103/PhysRevE.92.062715} {\bibfield  {journal} {\bibinfo
  {journal} {Phys. Rev. E}\ }\textbf {\bibinfo {volume} {94}},\ \bibinfo
  {pages} {022501} (\bibinfo {year} {2016})}\BibitemShut {NoStop}%
\bibitem [{\citenamefont {Suhonen}\ \emph {et~al.}(2014)\citenamefont
  {Suhonen}, \citenamefont {Kaski},\ and\ \citenamefont
  {Linna}}]{pauli_translocation}%
  \BibitemOpen
  \bibfield  {author} {\bibinfo {author} {\bibfnamefont {P.~M.}\ \bibnamefont
  {Suhonen}}, \bibinfo {author} {\bibfnamefont {K.}~\bibnamefont {Kaski}}, \
  and\ \bibinfo {author} {\bibfnamefont {R.~P.}\ \bibnamefont {Linna}},\
  }\bibfield  {title} {\enquote {\bibinfo {title} {Criteria for minimal model
  of driven polymer translocation},}\ }\href {\doibase
  10.1103/PhysRevE.90.042702} {\bibfield  {journal} {\bibinfo  {journal} {Phys.
  Rev. E}\ }\textbf {\bibinfo {volume} {90}},\ \bibinfo {pages} {042702}
  (\bibinfo {year} {2014})}\BibitemShut {NoStop}%
\bibitem [{\citenamefont {Cacciuto}\ and\ \citenamefont
  {Luijten}(2006{\natexlab{b}})}]{cacciuto_free_energy}%
  \BibitemOpen
  \bibfield  {author} {\bibinfo {author} {\bibfnamefont {A.}~\bibnamefont
  {Cacciuto}}\ and\ \bibinfo {author} {\bibfnamefont {E.}~\bibnamefont
  {Luijten}},\ }\bibfield  {title} {\enquote {\bibinfo {title} {Self-avoiding
  flexible polymers under spherical confinement},}\ }\href@noop {} {\bibfield
  {journal} {\bibinfo  {journal} {Nano Letters}\ }\textbf {\bibinfo {volume}
  {6}},\ \bibinfo {pages} {901} (\bibinfo {year}
  {2006}{\natexlab{b}})}\BibitemShut {NoStop}%
\bibitem [{\citenamefont {Gardiner}(2009)}]{gardiner_stochastic_methods}%
  \BibitemOpen
  \bibfield  {author} {\bibinfo {author} {\bibfnamefont {Crispin}\ \bibnamefont
  {Gardiner}},\ }\href@noop {} {\emph {\bibinfo {title} {Stochastic methods}}}\
  (\bibinfo  {publisher} {Springer Berlin},\ \bibinfo {year}
  {2009})\BibitemShut {NoStop}%
\bibitem [{\citenamefont {Al~Lawati}\ \emph {et~al.}(2013)\citenamefont
  {Al~Lawati}, \citenamefont {Ali},\ and\ \citenamefont
  {Al~Barwani}}]{lawati_capsid_tail}%
  \BibitemOpen
  \bibfield  {author} {\bibinfo {author} {\bibfnamefont {Afaf}\ \bibnamefont
  {Al~Lawati}}, \bibinfo {author} {\bibfnamefont {Issam}\ \bibnamefont {Ali}},
  \ and\ \bibinfo {author} {\bibfnamefont {Muataz}\ \bibnamefont
  {Al~Barwani}},\ }\bibfield  {title} {\enquote {\bibinfo {title} {Effect of
  temperature and capsid tail on the packing and ejection of viral {DNA}},}\
  }\href {\doibase 10.1371/journal.pone.0052958} {\bibfield  {journal}
  {\bibinfo  {journal} {PLOS ONE}\ }\textbf {\bibinfo {volume} {8}},\ \bibinfo
  {pages} {e52958} (\bibinfo {year} {2013})}\BibitemShut {NoStop}%
\end{thebibliography}%

\pagebreak

\end{document}